\documentclass[conference]{IEEEtran}
\IEEEoverridecommandlockouts

\usepackage{epsfig, subfigure, amsmath, amssymb}

\usepackage{amsthm, algorithm, algpseudocode}
\usepackage{epstopdf}
\usepackage{balance}
\usepackage{color}
\usepackage{url}
\usepackage{pifont}
\usepackage{paralist}
\usepackage{commath}
\usepackage{enumitem}
\usepackage{balance}
\usepackage{comment}
\usepackage{xcolor}
\usepackage{soul}
\usepackage{graphicx}
\usepackage{mwe}

\definecolor{red}{rgb}{1,0,0}

\definecolor{blue}{rgb}{0,0,1}

\newcommand{\eg}{{\it e.g.}, }
\newcommand{\ie}{{\it i.e.}, }

\begin{document}
%\title{False Data Injection Attacks against Contingency Analysis in Power Grids}
\title{Novel Attacks against Contingency Analysis in Power Grids}

\author{
\IEEEauthorblockN{Mohammad Ashiqur Rahman\IEEEauthorrefmark{1}\IEEEauthorrefmark{3}, Md Hasan Shahriar\IEEEauthorrefmark{1}\IEEEauthorrefmark{3}, Mohamadsaleh Jafari\IEEEauthorrefmark{1}, Rahat Masum\IEEEauthorrefmark{2}{\thanks{\IEEEauthorrefmark{3}Rahman and Shahriar are the co-first authors of this paper.}}}

\IEEEauthorblockA{\IEEEauthorrefmark{1}Department of Electrical and Computer Engineering, Florida International University, Miami, FL, USA\\
\IEEEauthorrefmark{2}Department of Computer Science, Tennessee Tech University, Cookeville, TN, USA\\
Email: marahman@fiu.edu, mshah068@fiu.edu, mjafari@fiu.edu, rmasum42@students.tntech.edu}}

%\IEEEspecialpapernotice{(Regular Paper)}

% make the title area
\maketitle

\begin{abstract}
Contingency Analysis (CA) is a core component of the Energy Management System (EMS) in the power grid. The goal of CA is to operate the power system in a secure manner by analyzing the system subject to a contingency (e.g., the outage of a transmission line or a power generator) to determine the set points that will allow system operation without violation of constraints. The analysis in CA is conducted based on the output from State Estimation (SE), another core EMS module. However, it is also shown that an adversary can alter certain power measurements to corrupt the system states estimated by SE without being detected. Such a corrupted estimation can severely skew the results of the contingency analysis as it will provide a fake model to deal with. In this research, we formally model necessary interdependency relationships and systematically analyze these novel attacks on the contingency analysis. In particular, this research focuses on Security Constrained Optimal Power Flow (SCOPF) that finds out the optimal economic dispatches considering a single line failure (based on the $n - 1$ contingency analysis) and transmission line capacities. The proposed model is implemented and solved to find out potential threat vectors (i.e., a set of measurements to be altered) that can evade CA so that the system will face overloading situation on one or more transmission lines when some specific contingencies happen. We demonstrate our formal model on an IEEE 14 bus system-based case study and verify the results with a standard PowerWorld model. We further evaluate the model with respect to various attacks and grid characteristics.
\end{abstract}

\begin{IEEEkeywords}
Power Grid; Security Constrained  Optimal Power Flow; Formal Model
\end{IEEEkeywords}

\IEEEpeerreviewmaketitle

%**********************************************
\section{Introduction}
\label{Sec:Introduction}
\vspace{-3pt}

%\textcolor{red}{Cyber technologies are increasingly used in smart power grids with the promise of providing larger capacity, higher efficiency, and more reliability. While this integration helps energy providers to offer smarter services, real time demand responses, and economic advantages, power grids also become vulnerable to cyber attacks, particularly cyber intrusion and false data injection, which can cause improper controls leading to serious damages including power outages and destruction of critical equipment~\cite{McDaniel09, Kundur10}. }

With the advances in various communication technologies and improved intelligence in physical control devices, Cyber-Physical Systems (CPSs) emerged as complex systems that can connect legacy control systems with the cyber world enabling two-way flow of information~\cite{McDaniel09, Kundur10}. Power grids are perfect examples of CPSs where cyber technologies are increasingly being used for smarter operations, economic advantages, and efficient management. However, these smarter grids are growingly becoming vulnerable to cyber attacks.

\begin{figure}[t]
%\vspace{-9pt}
\begin{center}
\includegraphics[scale=0.52,keepaspectratio=true]{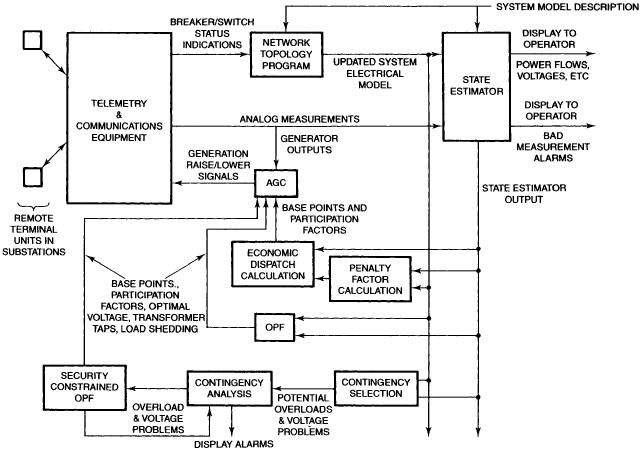}
\vspace{-9pt}
\caption{A shematic of Energy control center system security schematic~\cite{wood1996power}}
\label{Fig_EMS}
\end{center} 
\vspace{-12pt}
\end{figure}

In a power grid, Energy Management System (EMS) refers to a set of modules used by Supervisory Control and Data Acquisition (SCADA) system for system wide data analysis, control, and operation. A schematic diagram of EMS is shown in Fig.~\ref{Fig_EMS}~\cite{wood1996power}. State Estimation (SE) estimates the system state variables from a set of real-time  SCADA data (i.e., power measurements and breaker/switch statuses). The states are used to compute the unknown power measurements as well as to rectify the received measurements. As seen in Fig.~\ref{Fig_EMS}, the modules are interdependent, i.e., the output of one module, e.g., SE, is required by several other modules, including Security Constrained Optimal Power Flow (SCOPF), for economic dispatch calculation and security assessment. Therefore, a problem in one module cascades to other EMS operations, causing an ultimate failure.

It has been shown that stealthy attacks can negatively affect the performance of the state estimation solver, often named as Undetected False Data Injection (UFDI) attacks, where adversaries can corrupt the SE solution by injecting false data to the measurements while remaining undetected~\cite{Liu09}. The basis of these types of attacks is to alter the measurements intelligently considering being consistent with the power system balances. In SE, a bad data detection (BDD) mechanism is traditional which utilizes redundant measurements to filter out erroneous meter measurements by checking the difference between observed and estimated measurements~\cite{Monticelli99, Abur04}. An adversary with sufficient knowledge of the measurement model can manipulate measurements and bypass BDD~\cite{Liu09}. %In~\cite{Bobba10,Kim11}, it is shown that UFDI attacks by adversaries with perfect knowledge can be defended if a strategic set of measurements is secured (\ie data integrity protected).

%%%%%%%%%%%%%%%%%%%%%%
%%%%%%%%%%%%%%%%%%%%%%

This paper aims at understanding the effects of UFDI attacks on CA, focusing on SCOPF. As shown in Fig.~\ref{Fig_EMS}, SCOPF determines the set-points of the generators for Automatic Generation Control (AGC). SCOPF allocates the best set-points by minimizing the generation cost while satisfying the system constraints. One of the most important constraints is the safety of the transmission lines when any line trips by conducting CA.  
%Thus, by calculating SCOPF, the system finds the generator set-points which ensure the minimum generation cost while securing the system from any contingency. These set-points are utilized by AGCs to regulate the generators' power output. 
% SCOPF calculation is mostly based on state estimation solver outputs and power grid topology. Switches and circuit breakers across the power grid send their status (open or closed) to the topology processor. Then, topology processor specify the grid topology based on these statuses. Fig.~\ref{Fig_EMS}, shows that various procedures in EMS including state estimation and SCOPF, are performed based on the data received from the measurements in the power system. 

The measurements are sent from the field devices through different communication channels to SE for estimating the whole grid. False measurements can mislead the estimation and so the SCOPF solution. As a result, no longer secured for the system. 
For instance, SCOPF can select an insecure set of generation which will run without any constraint violation but may overload some transmission lines if any contingency happens. However, these attacks must bypass BDD to remain stealthy, while the attacker cannot exceeds its capability. Exploring these attack vectors is an important challenge. 
Considering this prime challenge, we introduce a framework capable of systematically analyzing the influence of UFDI attacks on CA/SCOPF of a power grid. Our contributions in this paper are as follows:

\begin{itemize}
\item We propose a formal framework to analyse the impact of false data injection attacks on SCOPF. This framework consists of modeling of the power system including SE and SCOPF, UFDI attacks, adversaries capabilities, and impact on SCOPF. 
In FDI attacks, line and bus power measurements are attacked to introduce changed power consumption at different buses. Based on wrong consumption data, the system calculates the SCOPF and run accordingly which is not optimal in reality. 

\item A Satisfiability Modulo Theories (SMT) solver is used to implement the proposed framework. We model this framework as a constraint satisfaction problem~\cite{Z3}. We evaluate the implemented tool and analyze the threat space according to different attack models. 
%In the following section, an example is given to illustrate the attack while the way that this framework verifies the possibility of stealthy attacks for an intended impact on SCOPF. Since our framework includes topology attacks and SCOPF and these systems mostly work with real values, we apply several strategies to increase the scalability.  
%the model complexity is high. Therefore, we analyze the model's complexity and propose mechanisms to improve its efficiency in terms of time and memory requirements.
\end{itemize}

The remaining parts of this paper are organized as follows:  Section~\ref{Sec:Background} presents background information and introduces an example of the attack model. Section~\ref{Sec:Model} contains formalization of attacks and their impact on SCOPF. Evaluation results are discussed in Section~\ref{Sec:Evaluation}. The related work is concisely reviewed in Section~\ref{Sec:Related}. Conclusion is presented in \ref{Sec:Conclusion}.

%***********************************************************
\section{Background and Motivation}
\label{Sec:Background}
%***********************************************************
%\section{Background and Motivation}
%\label{Sec:Background}
% Background, Related Work, and Motivation here

Here we discuss necessary background information related to this research. 

%%%%%%%%%%%%%%%%%%%%%%%%%%%%%%%%
\subsection{DC Power Flow}
\label{SubSec:DCPowerFlow}

The most common and important calculation for a power system is the AC power flow analysis. However, an AC power flow analysis is computationally expensive as a grid with $n$ nodes need to solve $2n$ non-linear equations through iteration. The high level of accuracy with a detailed result provided by the AC power flow model does not overcome the high computational expense. Therefore, a linearized model of the system is used to speed up the computation which is called  DC power flow model~\cite{purchala2005usefulness}. To make the model less expensive, some assumptions are made which speed up the computation with the trade of accuracy. However, the result of the DC power model is accurate enough the load flow calculation. The DC power flow model is discussed further in Appendix~\ref{Appendix:DCPowerFlowModel} for the interested readers.
\subsection{State Estimation and Stealthy Attacks}
%\vspace{-3pt}

State estimation of DC power flow model estimates the phase angles of the nodes from the several transmission line power flow measurements.  Specially, to estimate $n$ number of states variables  $\mathbf{x} = (x_1, x_2, \cdots , x_n)^T$, $m$ number of meter measurements $\mathbf{z} = (z_1, z_2, \cdots , z_m)^T$are used~\cite{Abur04}.  
As for the assumptions of the DC power flow model discussed in section \ref{SubSec:DCPowerFlow}, 
the measurement model is linear 
%(\ie the measured power-flows are linear functions of bus phase angles) 
and hence the measurement model can be written as:
\vspace{-3pt}
$$\mathbf{z} = \mathbf{H} \mathbf{x} + \mathbf{e} \textrm{, where } \mathbf{H} = (h_{i, j})_{m \times n}$$ 

While $n$ number of measurements are sufficient to calculate the states, usually $m >> n$, which indicates that the measurement set includes redundant measurements. 
%, \ie $m > n$ which comprises an over-determined set of linear equations. 
The gross measurement errors are detected, removed and smoothed by the redundancy. Considering Gaussian distribution for the the measurement error with zero mean and known deviation, the state estimation $\hat{\mathbf{x}}$ is given as:
\vspace{-3pt}
\begin{equation}
\hat{\mathbf{x}} = (\mathbf{H}^T \mathbf{W} \mathbf{H})^{-1} \mathbf{H}^T \mathbf{W} \mathbf{z} \label{statesol}
\end{equation}
Here, $\mathbf{W} $ is a diagonal  ``weighting" matrix whose elements are reciprocals of variances of the meter errors. Thus, the measurement values can be calculated by $\mathbf{H\hat{x}}$  when  $||\mathbf{z} - \mathbf{H}\hat{\mathbf{x}}||$ provides the measurement residual.
%which would be used to determine bad data. 
Considering $\tau$ as the threshold for the errors, any data satisfying $||\mathbf{z} - \mathbf{H}\hat{\mathbf{x}}|| > \tau$  is detected as a bad data.

While injecting false data into the system, the bad data detection process can be bypassed by generating attack vectors accordingly~\cite{Liu09}. A random value of false data $\mathbf{a}$ can be injected to the measurements $\mathbf{z}$ so that $\mathbf{a}$ is a linear combination of the column vectors of $\mathbf{H}$ such as $\mathbf{a} = \mathbf{H}\mathbf{c}$ where $\mathbf{c}$ is added to the original state estimate $\hat{\mathbf{x}}$ to avoid the injection of $\mathbf{a}$ into the measurements. Since $\mathbf{z} + \mathbf{a} = \mathbf{H}(\hat{\mathbf{x}} + \mathbf{c})$, the residual $||(\mathbf{z} + \mathbf{a}) - \mathbf{H} (\hat{\mathbf{x}} + \mathbf{c})||$ still remains the same as $||\mathbf{z - H\hat{x}}||$ and the bad data injection is not detected. Note that this requires knowledge of $\mathbf{H}$, \ie full knowledge of system topology, parameters and measurement configuration.
The same approach is used to bypass the UFDI into the system to attack the contingency analysis of the power grids.

\subsection{Security Constrained Optimal Power Flow}
%\vspace{-3pt}

The goal of the SCOPF algorithm is to minimize the total generation cost by adjusting different system controls while satisfying power balance constraints imposing the elements' maximum rated capacities.  However, the secure and reliable operation of the power system expects no uncontrollable situations during the contingencies. In order to make the system secure, the optimization of the cost function is needed to consider the contingency scenarios as an additional constraint. This procedure is called SCOPF algorithm. The SCOPF algorithm adjusts the control settings during the pre-contingency condition in order to avoid interruption in the post-contingency conditions. 

Consequently, SCOPF algorithm intends to minimize  the total cost of {\em generation} satisfying the constraints: (i) the equality of total power generation and consumption (ii) the equipment ratings, transmission line capacity limits and control variables both in pre-contingency and post-contingency conditions ~\cite{Wood96}.  Considering the cost of generation of generator $k$ by $C_k(P_k^G)$, where $C_k$ has two parts: fixed cost and variable cost both depending on the nature of plant (\eg type of fuel, type of machine, type of cycle, etc.).

Thus, the SCOPF algorithm (with the DC flow model) can be described as follows:
\vspace{-3pt}
\begin{eqnarray}
 \textbf{Power~Flow~Constraint:}~~\mathbf{[B] } \mathbf{[\theta]} = \mathbf{[P]}  \label{powerflow} \\
\textbf{Line~Flow~Limit~(pre-contingency):}~~ | P_{i} | \leq P_{i}^{max} \label{lineflow} \\
\textbf{Line~Flow~Limit~(post-contingency):}~~ | P^{}_{(i,j)} | \leq P_{i}^{max} \label{lineflowca} \\
\textbf{Equipment~Rating:~~}P_{k}^{min}  \leq P_{k} \leq P_{k}^{max} \label{gencon}\\
\textbf{Optimization~Goal:~~}min \sum_{k} C_k(P_k) \textrm{ s.t.}  \label{obj} 
\end{eqnarray}
%
%Here Equation~\ref{powerflow} represents the power-flow-analysis constraint, power flow limit for line $i$ during the pre-contingency period is represented by Equation \ref{lineflow}, 
Here, the power flow limit for line $i$ when line $j$ is tripped is shown in Equation ~\ref{lineflowca}, equipment (\ie generator, transformer)  rating constraint is shown in Equation~\ref{gencon}. Finally, the optimization target is to minimize the cost of operation (Equation~\ref{obj}). 

%We next characterize attacks in terms of attributes to assess their impact  on the OPF.  

%\vspace{-3pt}
%%%%%%%%%%%%%

\subsection{Attack Model}
\vspace{-3pt}

In this work, cyber attacks on the base of FDI is considered in order to assess their impact on SCOPF. The false data can be injected by compromising the corresponding sensor or meter during the transmission of the data, as a man-in-the-middle attack. The sensors are located at the bus or the remote terminal unit (RTU). 

%It is worth mentioning that measurements within a substation are often gathered at a single RTU. that an adversary can break into to access the measurement data

%Here, we characterize attacks in terms of attributes to assess their impact on the SCOPF. Our goal is to describe attacks in their most general form so that adversarial capabilities can be modeled. 
%This also allows us to determine under what conditions, a given attack is feasible. 
The parameters of the attack model mainly consists of the attacker's resources, accessibility, and being informed of the system.
%
%\noindent\emph{Attacker's Accessibility and Resources}: 
%
%\textit{Accessibility}: 
Physical or remote access to some of the measurements can be restricted to the attacker. Therefore, to inject the false data, the attacker should be able to access the substation or the corresponding RTU. Some of the measurements may be secured. An adversary also needs to have necessary knowledge of the grid, in particular the bus topology and the transmission lines' electrical properties (e.g., impedances).
%
%\textit{Resource Constraint}: 
Furthermore, an adversary might be limited in terms of cost or difficulties of launching attacks on highly scattered measurements. It is harder for an attacker to inject false data to highly scattered measurements than measurements distributed in a less number of substations. 

\vspace{-3pt}

\subsection{An Example of FDI Attack on SCOPF}
\begin{figure}[t]
%\vspace{-9pt}
\begin{center}
\includegraphics[scale=0.35,keepaspectratio=true]{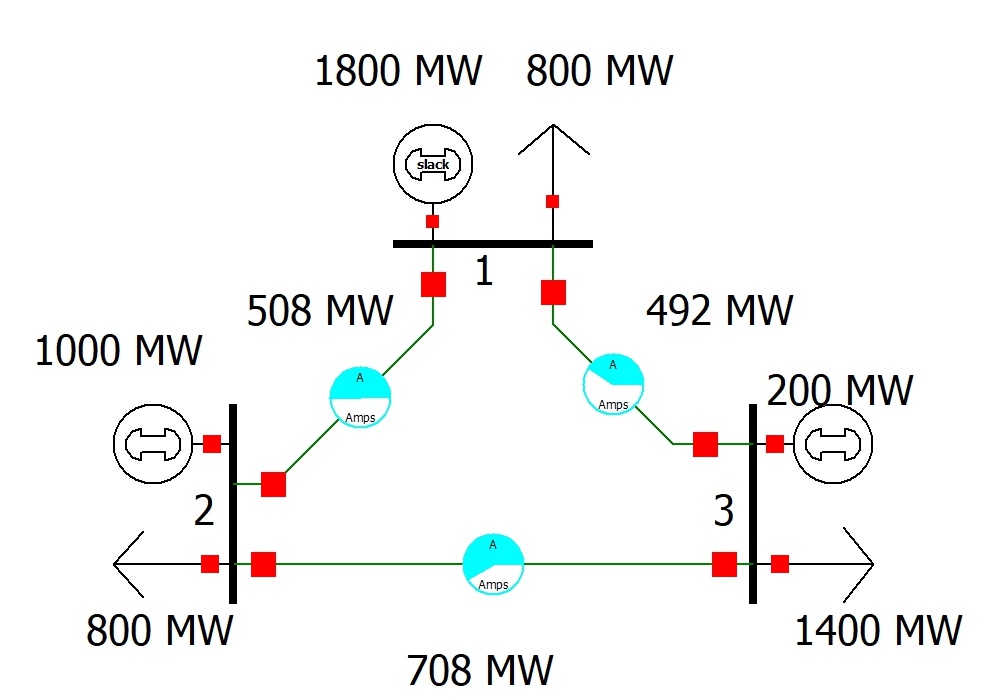}
\caption{3 bus power system}
\label{Fig_3busSys}
\end{center} 
\vspace{-6pt}
\end{figure}

%%%%%%%%%%%%%%%%%%%%%%%%%%%%%%%%%%%%%%%%%%%%%%%
To understand the FDI attack on SCOPF, consider a simple 3 bus power system where all the buses have both generator and load and are interconnected to each other through transmission lines as shown in Fig. \ref{Fig_3busSys}. Table~\ref{3busTable} shows the bus and line parameters. The rated loads of buses 1, 2, and 3 are 1000 MW, 1000 MW, and 1500 MW, respectively, while their current loads are 800 MW, 800 MW, and 1400 MW, respectively. The rated generation of the buses is 2000 MW, 1500 MW, and 500 MW, respectively. The cost coefficients pairs of the generators are (10, 1), (10, 2), and (10,3) respectively. Line 1 connects bus 1 and 2, line 2 connects bus 1 and 3 and line 3 connect bus 2 and 3. The rated capacities of the transmission lines are 1100 MW, 1200 MW, and 1200 MW, respectively. Thus, in the pre-attack (no attack) condition, the SCOPF dispatches are 1800 MW, 1000 MW, and 200 MW, respectively with a generation cost of \$4,430. The line flows without any contingency are 508 MW, 492 MW, and 708 MW, respectively.

%%%%%%%%%%%%%%%%%%%%%%%%%%%%%%%%%%%%%%%%%%%%%%%%%%%%%%%%%%%%%%%%%%%%%%%

%%%%%%%%%%%%%%%%%%%%%%%%%%%%%%%%%%%%%%%%%%%%%%%%%%%%%%%%%%%%%%%%%%%%%%%%

\begin{figure*}[t]
    \begin{center}
    %\hspace{-12pt}            
        \subfigure[]{
            \label{preattack0}
            \includegraphics[scale=0.20, keepaspectratio=true]{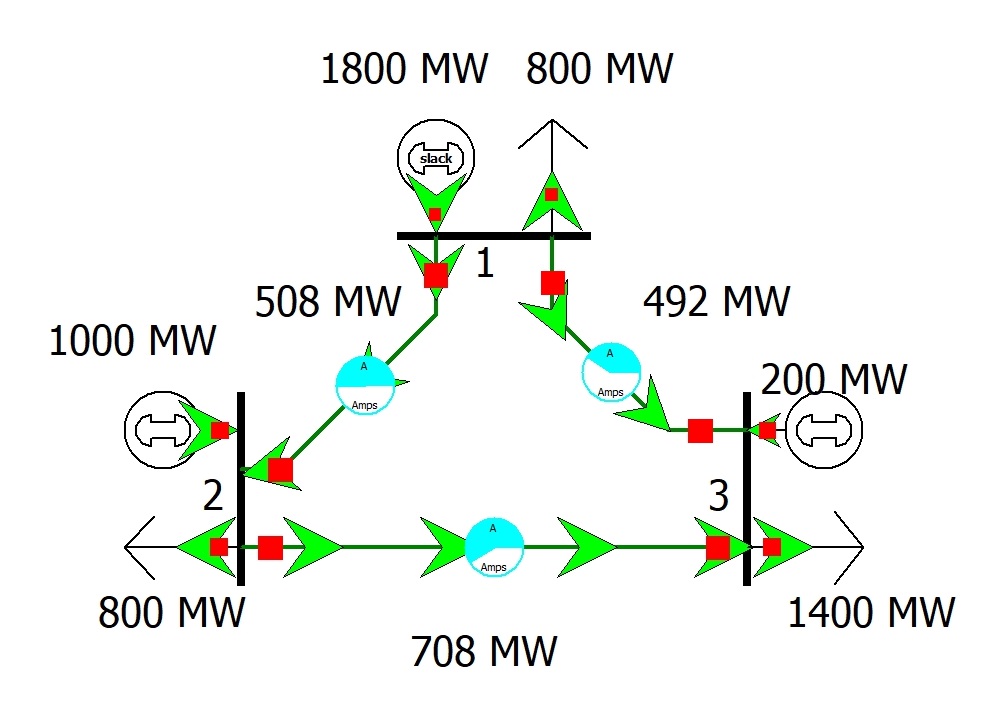}
        }   
        \subfigure[]{
            \label{preattack1}
            \includegraphics[scale=0.20, keepaspectratio=true]{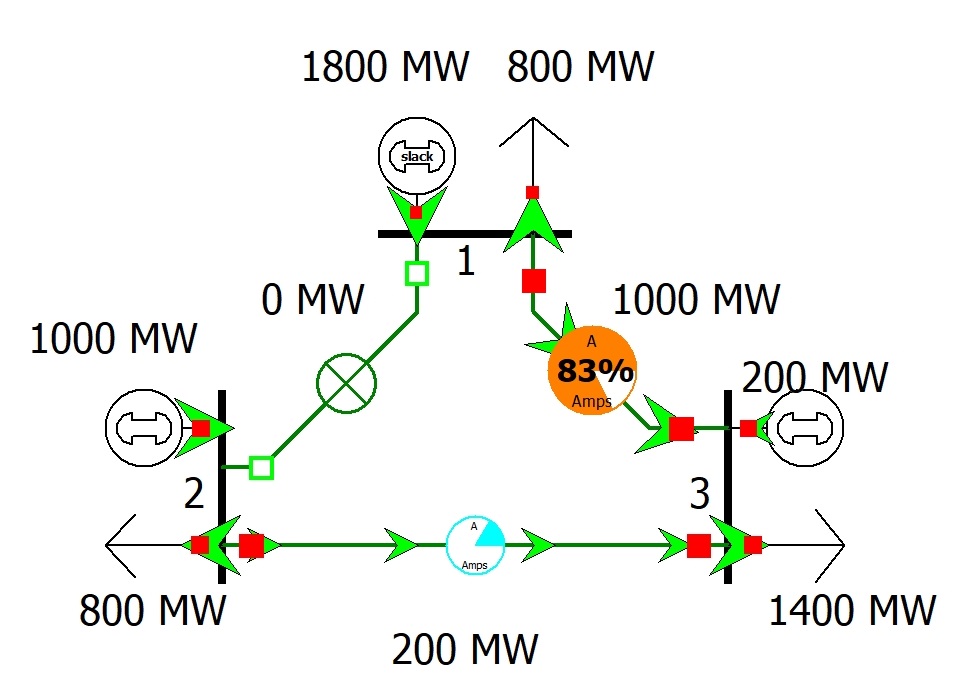}
        }
    \subfigure[]{
            \label{preattack2}
            \includegraphics[scale=0.20, keepaspectratio=true]{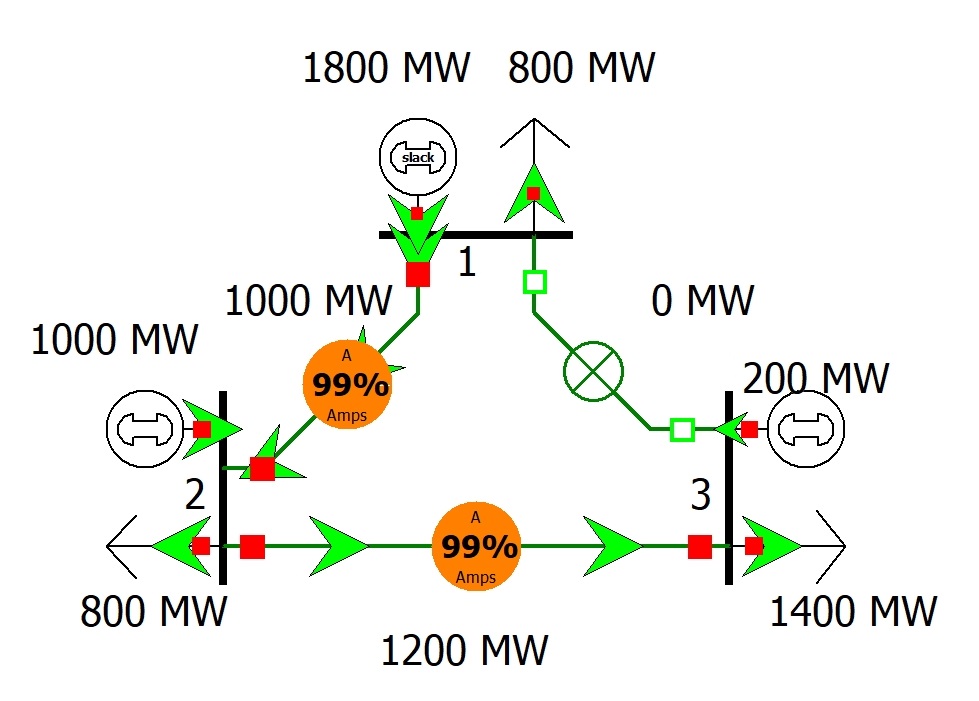}
        }
    \subfigure[]{
            \label{preattack3}
            \includegraphics[scale=0.20, keepaspectratio=true]{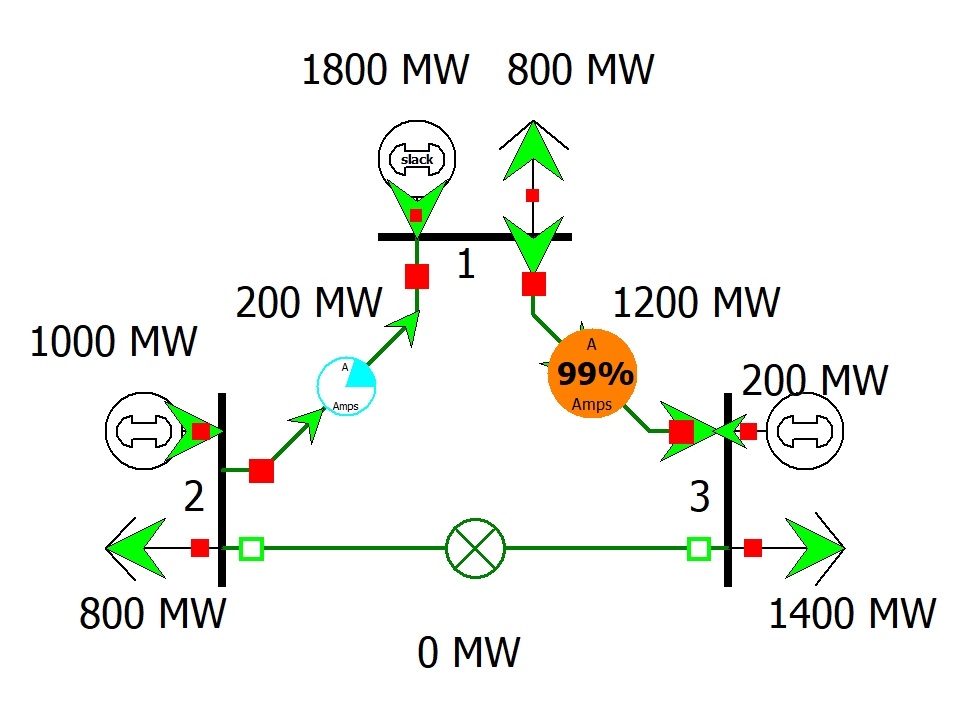}
        }

    \end{center}
    \vspace{-15pt}
    \caption{Contingency scenarios of pre-attack condition : (a) no contingency, (b) line 1 trips, and (c) line 2 trips (d) line 3 trips.}
    \label{Pre-attack}
%\vspace{12pt}
\end{figure*}

As the system running on the SCOPF solution, if any line trips the system should sustain and continue the operation. The three contingency conditions when line 1, 2, or 3 trips are shown in Fig.~\ref{Pre-attack}. From the figure, we see that all the line flows are within the threshold capacity limit, which ensures that the system sustains even when any line trips.

%%%%%%%%%%%%%%%%%%%%%%%%%%%%%%%%%%%%%%%%%%%%%%%%%%%%%%%%%%%%%%%%%%%%%%%%%

\begin{table}[t]
\vspace{-6pt}
\caption{Bus and line parameters of 3 bus power system} \label{3busTable}
\vspace{-6pt}
\centering
\scriptsize

\begin{tabular}{|l|l|l|l|}
\hline
\textbf{Bus Data}                    & \textbf{Bus 1}  & \textbf{Bus 2}  & \textbf{Bus 3}  \\ \hline
{Rated Load (MW)}             & 1000            & 1000            & 1500            \\ \hline
{Current Load (MW)}           & 800             & 800             & 1400            \\ \hline
{Generation Capacity (MW)}    & 2000            & 1500            & 500             \\ \hline
{Cost Coefficients (\$, \$/MW)} & (10, 1)         & (10, 2)         & (10, 3)         \\ \hline
{SCOPF Dispatch (MW)}         & 1800            & 1000            & 200             \\ \hline
\multicolumn{4}{|l|}{}                                                                     \\ \hline
\textbf{Line Data}                   & \textbf{Line 1} & \textbf{Line 2} & \textbf{Line 3} \\ \hline
{Connecting Buses}            & 1, 2            & 1, 3            & 2, 3            \\ \hline
{Rated Capacity (MW)}         & 1100            & 1200            & 1200            \\ \hline
{Current Line Flow (MW)}      & 508             & 492             & 708             \\ \hline
{Percent of overloading (\%)} & 46              & 41              & 59              \\ \hline
\end{tabular}
\vspace{-6pt}
\end{table}
%%%%%%%%%%%%%%%%%%%%%%%%%%%%%%%%%%%%%%%%%%%%%%%%%%%%%%%%%%%%%%%%%%%%%%%
%------------------ Pre Attack -------------
%---------------------------------------------------
%------------------ Post Attack : Expected ---------
\begin{figure*}[t]
    \begin{center}
    %\hspace{-12pt}            
        \subfigure[]{
            \label{exattack0}
            \includegraphics[scale=0.20, keepaspectratio=true]{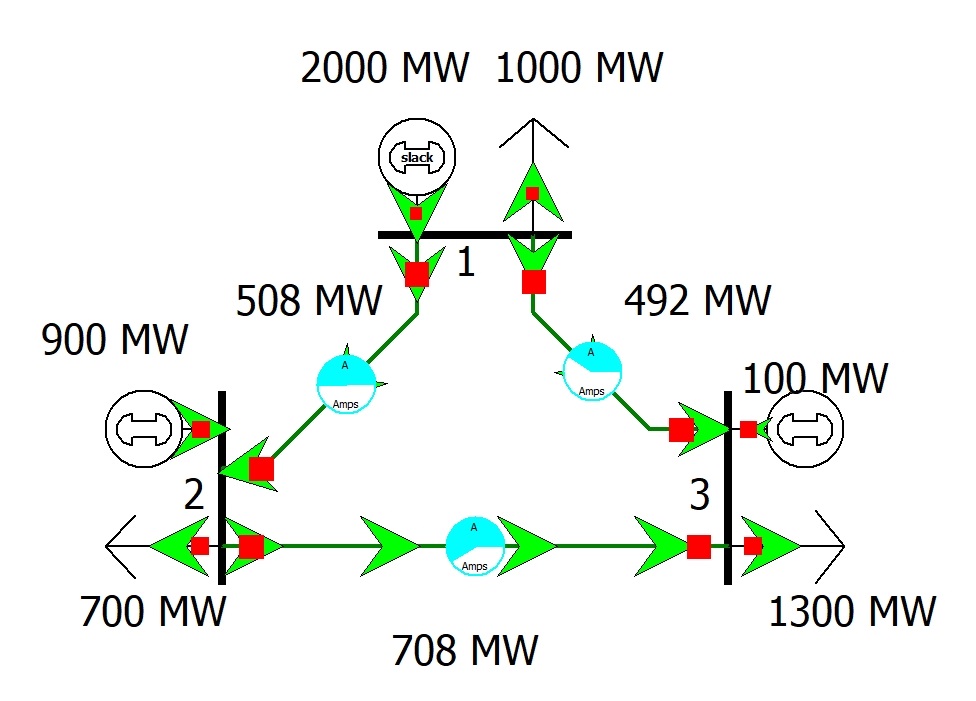}
        }   
        \subfigure[]{
            \label{exattack1}
            \includegraphics[scale=0.20, keepaspectratio=true]{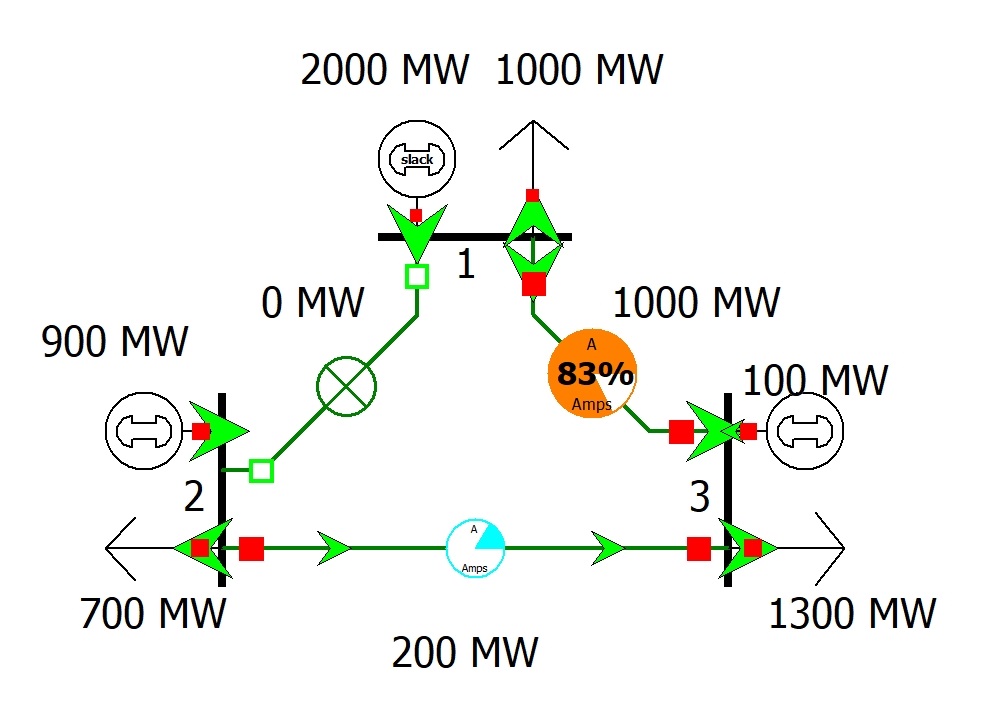}
        }
    \subfigure[]{
            \label{exattack2}
            \includegraphics[scale=0.20, keepaspectratio=true]{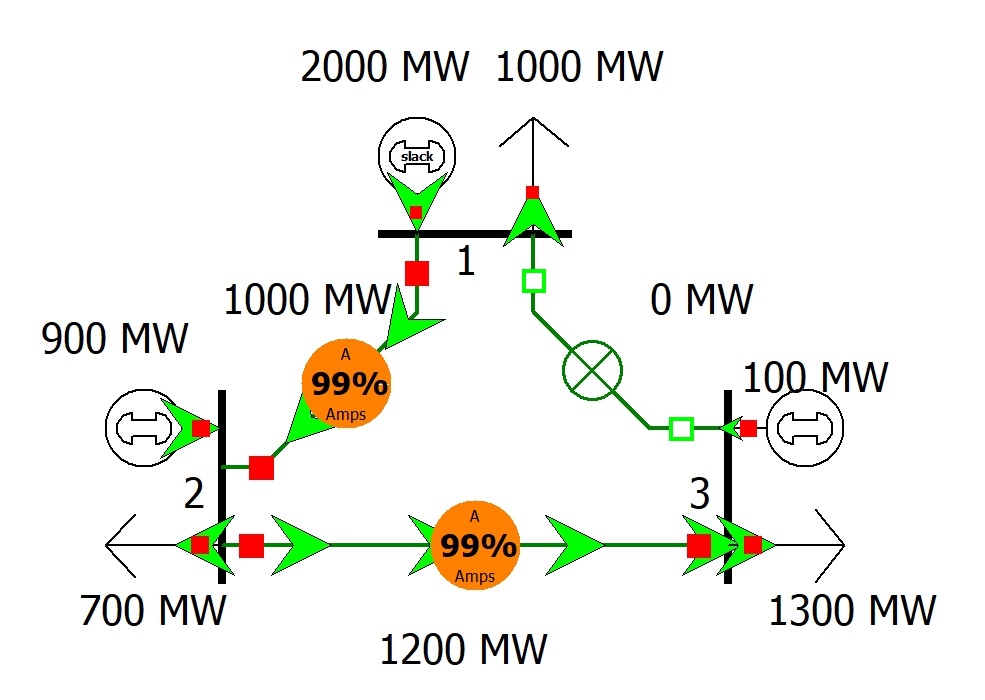}
        }
    \subfigure[]{
            \label{exattack4}
            \includegraphics[scale=0.20, keepaspectratio=true]{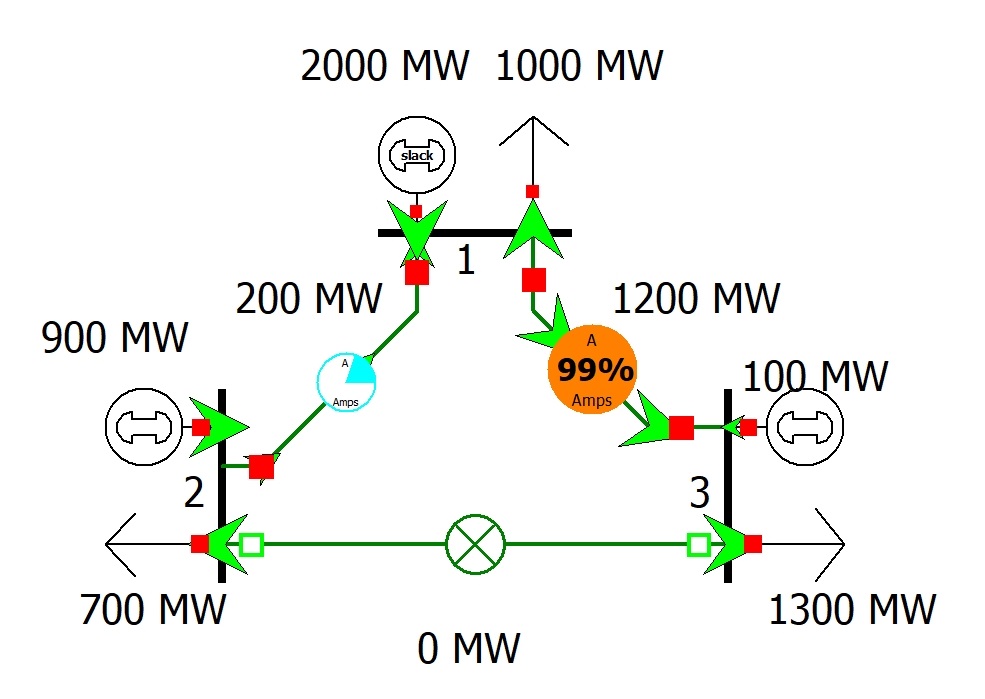}
        }

    \end{center}
    \vspace{-15pt}
    \caption{Expected contingency scenarios after the attack: (a) no contingency, (b) line 1 trips, and (c) line 2 trips (d) line 3 trips.}
    \label{Exp_post_attack}
%\vspace{12pt}
\end{figure*}
%---------------------------------------------------
%------------------ Post Attack : Expected ---------

\begin{figure*}[t]
    \begin{center}
    %\hspace{-12pt}            
        \subfigure[]{
            \label{reattack0}
            \includegraphics[scale=0.20, keepaspectratio=true]{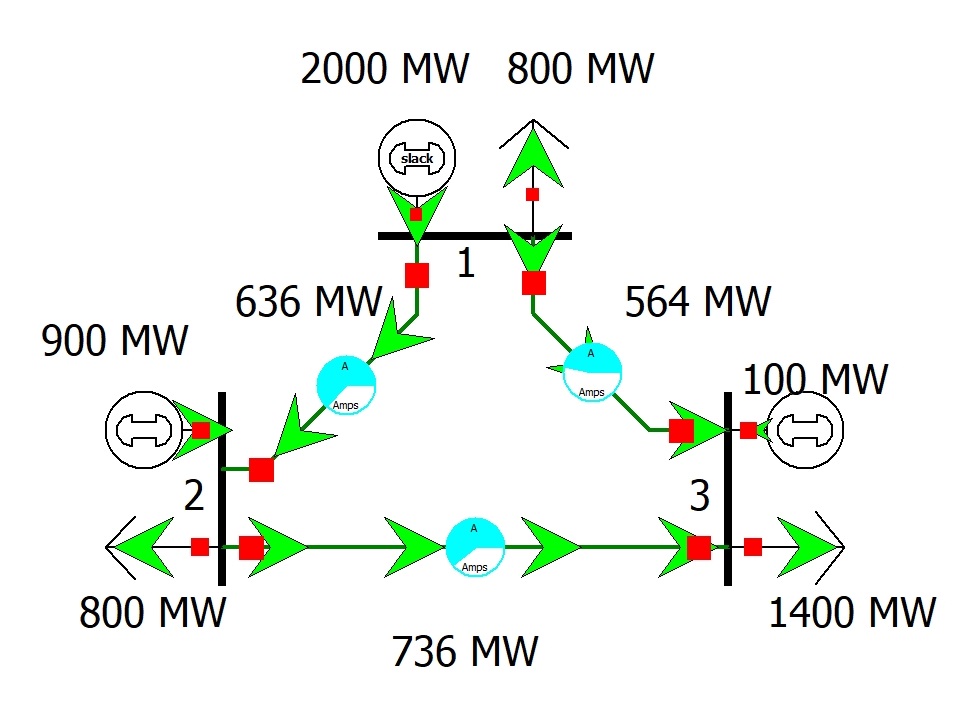}
        }   
        \subfigure[]{
            \label{reattack1}
            \includegraphics[scale=0.20, keepaspectratio=true]{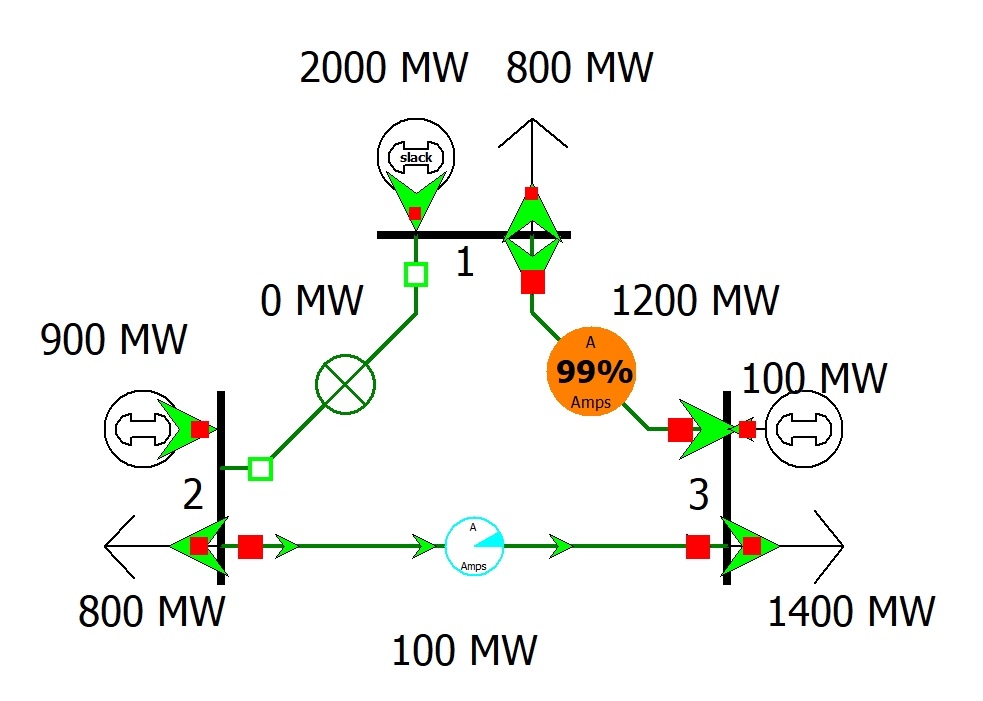}
        }
    \subfigure[]{
            \label{reattack2}
            \includegraphics[scale=0.20, keepaspectratio=true]{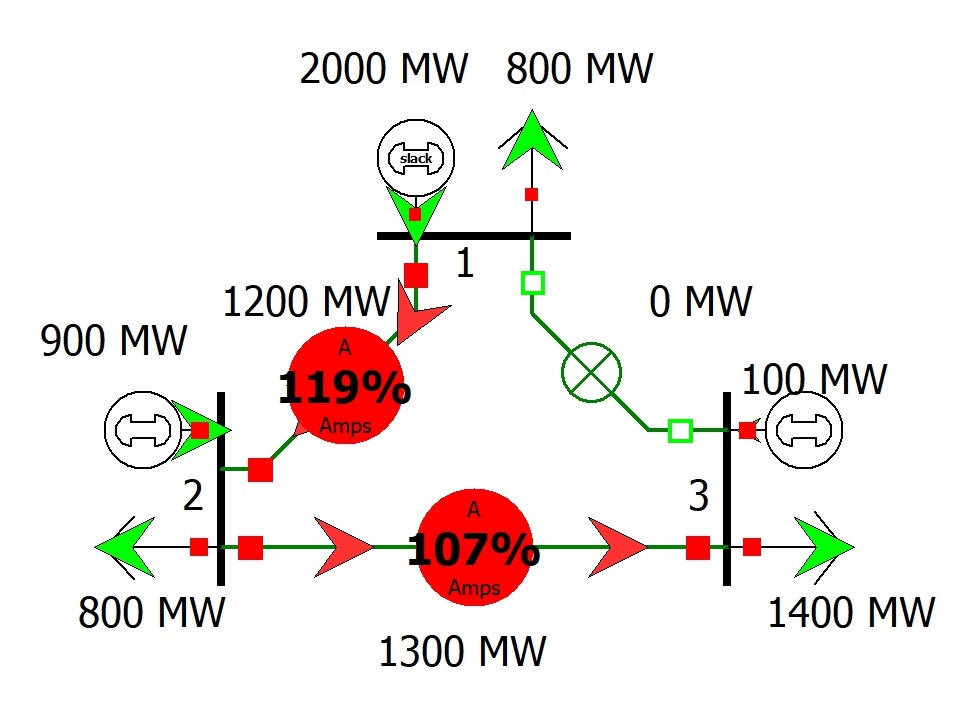}
        }
    \subfigure[]{
            \label{reattack3}
            \includegraphics[scale=0.20, keepaspectratio=true]{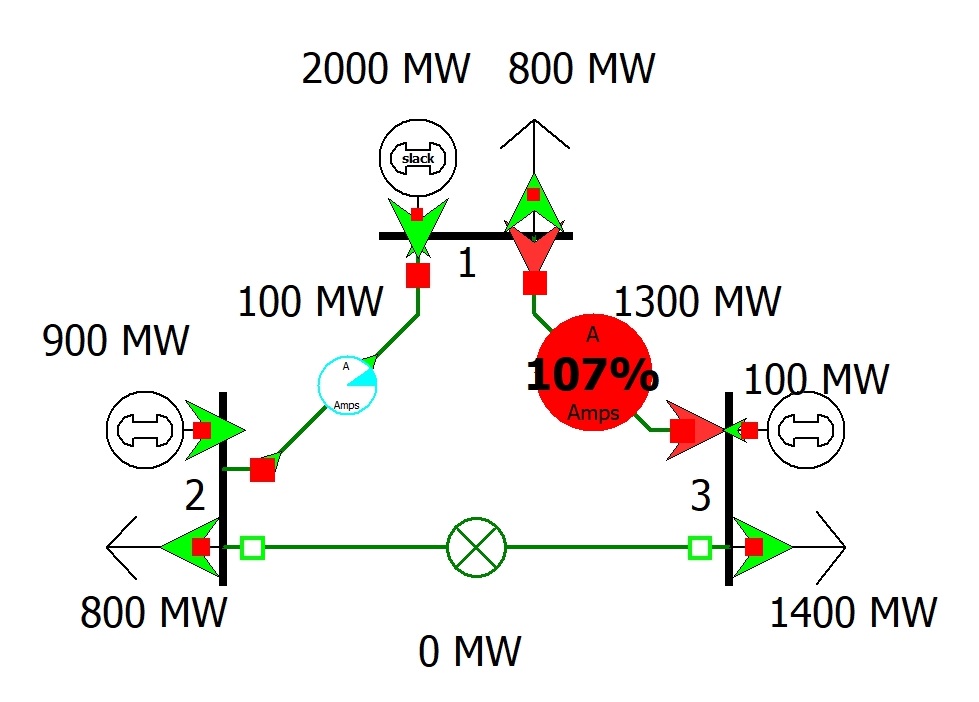}
        }

    \end{center}
    \vspace{-15pt}
    \caption{Post attack contingency scenarios with real data: (a) no contingency, (b) line 1 trips, and (c) line 2 trips (d) line 3 trips.}
    \label{Real_post_attack}
%\vspace{12pt}
\end{figure*}
%---------------------------------------------------
%%%%%%%%%%%%%%%%%%%%%%%%%%%%%%%%%%%%%%%%%%%%%%%%%%%%%%%%%%%%%%%%%%%%%%%
%%%%%%%%%%%%%%%%%%%%%%%%%%%%%%%%%%%%%%%%%%%%%%%

At this point, the attacker injects some false data into the system in a way that the EMS observes 200 MW of load increase in bus 1. In fact, it is a load shift/redistribution from buses 2 and 3 to bus 1. Thus, the system now expects 1000 MW in bus 1, 700 MW in bus 2, and 1300 MW in bus 3. Based on the false load data, the system will calculate the new set of generations following the SCOPF algorithm. Thus, after running the SCOPF on the false data, the new generation dispatches are 2000 MW, 900 MW, and 100 MW, respectively. Again with the attacked false data, the system expects that all line flows will be placed within the capacity limit during normal and contingency conditions. Fig.~\ref{Exp_post_attack} shows the different scenarios expected by EMS during the contingencies. With the new generation dispatches, the total cost is \$4,130 which is lower than the pre-attack cost. 

However, the actual values are still the same as the pre-attack condition, misguiding the system to run on another set point which is not optimal anymore. The real loads are still the same but the generation dispatches are different from the optimum values. Thus, the system will run at a different condition than the expected one. Fig.\ref{Real_post_attack} shows the scenarios that the system observes during normal and contingency conditions. It can be seen that, though during the normal operation of the system there is no line overloading if one of the lines is tripped. It is shown that tripping of line 1 does not create any overloading among other lines. However, when line 2 trips, line 1 and line 3 get overloaded by 109\% and 107\%. Moreover, when line 3 trips, line 2 becomes overloaded by 107\% where the EMS is expecting no overloading.
\section{Formal Model for Synthesizing UFDI Attacks that Impacts SCOPF}
\label{Sec:Model}
%\vspace{-3pt}

In this section, we first briefly discuss the  basic idea/framework of verifying the impact of stealthy attacks on SCOPF. Then, we discuss the corresponding formalization in detail. Finally, we demonstrate the execution of the proposed formal framework on some example cases. 

%%%%%%%%%%%%
\subsection{Framework}
\label{SubSec:Model_Framework}

\begin{figure}[t]
\begin{center}
\vspace{-12pt}
\includegraphics[width=\columnwidth]{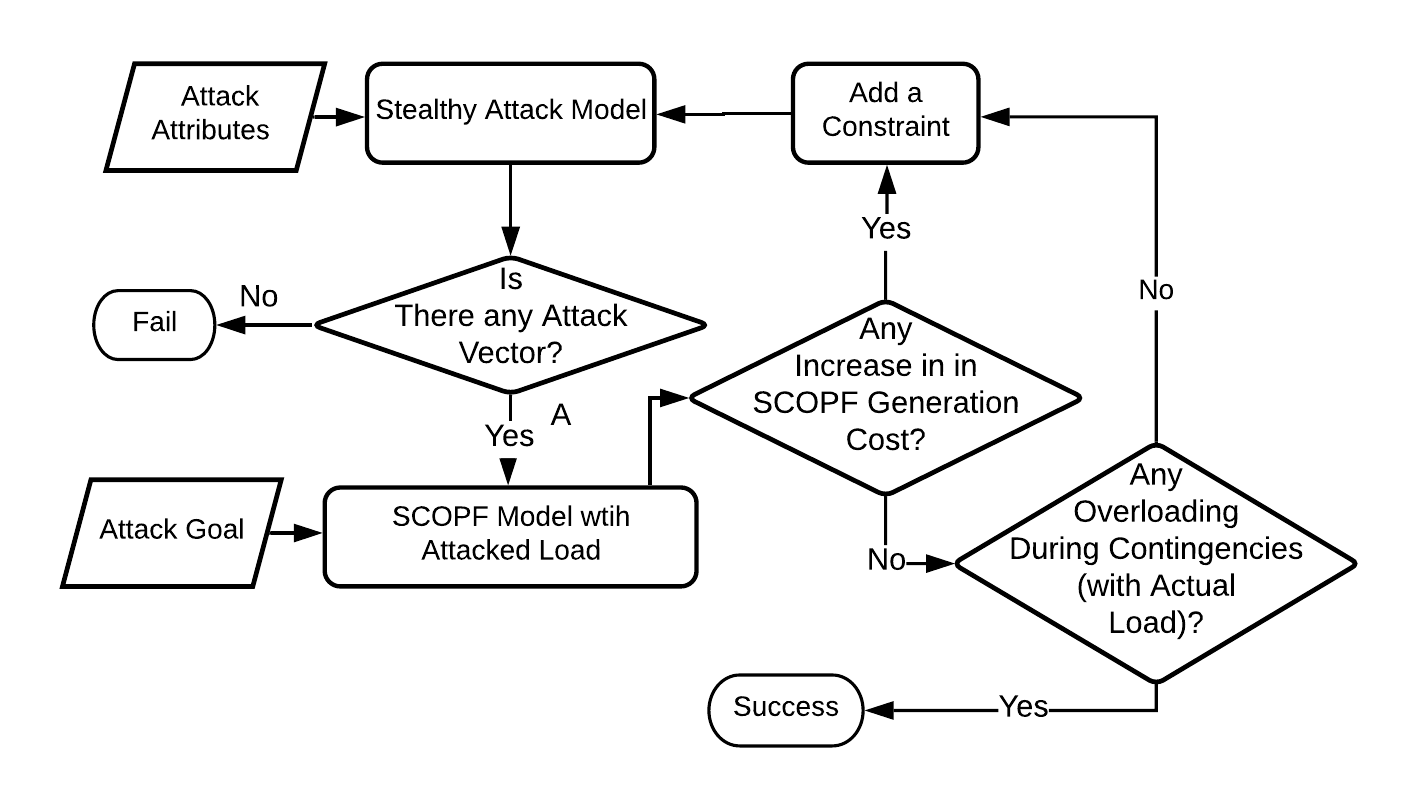}
\vspace{-6pt}
\caption{The systematic flow of finding the impact of UFDI attacks on SCOPF. The proposed formal model unifies this concept in a single  model.}
\label{Fig_Flow_Diagram}

\end{center}
\vspace{-15pt}
\end{figure}

The synthesis of a stealthy attack that impact/compromise CA/SCOPF follows the idea presented in Fig.~\ref{Fig_Flow_Diagram}. The idea is composed of two basic formal model: \emph{UFDI attack model} that finds potential attack vectors corresponding and \emph{SCOPF model} that verifies if there exists an SCOPF solution within a threshold cost. The objective is to launch an stealthy attack on CA without increasing the generation cost.

%%%%%%%%%%%%%%%%%%%%%%%
%%%%%%%%%%%%%%%%%%%%%%%

Impact analysis includes a couple of stages: First, we try to find an attack vector that satisfies the attack model (\ie attack attributes). When there is an attack vector, the system is (incorrectly) updated according to the attack vector, \ie with the consideration of the changed loads. 
%% Shahriar Check [Start]
Then, the generation cost is evaluated by executing the SCOPF model to see if there is any increase in the generation cost with respect to the original (\ie no-attack) scenario. 
% I commented the following sentence [Ashiq]
%In order to verify this increase, the expected raise is added to the threshold original (\ie in the no-attack scenario) SCOPF solution set the threshold cost and check whether there is still an SCOPF solution within this threshold.
If no, then the attack was successful. Otherwise, the same steps will be taken to find a new attack vector until either we find an attack vector or no more attack vectors exist.
%% Shahriar check [End]
As mentioned earlier, the objective is to overload some lines during contingency while the generation cost remains unchanged (\ie no significant cost difference). % but not to fail the OPF process to converge considering the power generation limit of each generator and the capacity of each transmission line.
%(\ie no OPF solution is possible considering the power generation limit of each generator and the capacity of each transmission line). 
%
%Although the SCOPF model and the topology attack model can be executed separately the framework creates a single model by combining them as shown in Fig.~\ref{Fig_Flow_Diagram}.
The formal model presented in the following subsection unifies the conceptual building blocks (as shown in  Fig.~\ref{Fig_Flow_Diagram}) into a single  model. 

%The two models of this framework are combined together into a single model, although they can be executed separately as shown in Fig.~\ref{Fig_Flow_Diagram}.
%In the following, we describe the formalization of this framework in detail. % with these two models.

%\vspace{-3pt}
%%%%%%%%%%%%%
\subsection{Formalization}
%\subsection{Parameters}
\label{SubSec:Model_SCOPF}
%\vspace{-3pt}

The notations used for UFDI attack modeling on contingency analysis are defined in Table~\ref{Tab_Notation}.
%These parameters are shown in Table~\ref{Tab_Notation}. In the following, we describe them briefly. 

%%%%%%%%
\begin{table}[t]
\vspace{-12pt}
\caption{Modeling Parameters} \label{Tab_Notation}
\vspace{-6pt}
\centering
%\scriptsize
\begin{tabular}{|p{0.37in}|p{2.8in}|}
\hline
\textbf{Notation}    & \textbf{Definition} \\
\hline
$b$ & Bus numbers in the grid. \\
$l$ & Line numbers in the grid topology. \\
$f_i$ & \emph{from-bus} of line $i$. \\
$e_i$ & \emph{to-bus} of line $i$. \\
$d_i$ &  Line $i$ admittance. \\
%$g_i$ & Whether line $i$ admittance is known. \\
$P^L_i$ & Line $i$ power flow. \\
$P^B_j$ & Bus $j$ power consumption. \\
$\theta_j$ & The state value, \ie the voltage phase angle at bus $j$. \\
$n$ & Number of states. \\
$m$ & Number of possible measurements. \\
%$a_i$ & Whether measurement $z_i$ is required to be altered for the attack. \\
$a_i$ & Whether measurement $i$ is required to be altered for the attack. \\
%$c_j$ & Whether state $x_j$ is affected due to false data injection. \\
$c_j$ & Whether state $j$ is corrupted due to false data injection. \\
$h_j$ & Whether any measurement collected at bus $j$ is needs an alteration for the attack. \\
$t_i$ & If potential measurement $i$ is taken (\ie recorded/reported by a meter). \\
$r_i$ & Whether the attacker can access measurement $i$. \\
$s_i$ & Whether measurement $i$ is secured. \\
$\delta_b$ & maximum false data of the as bus load.\\
$\delta_l$ & minimum overloading of the lines during contingencies.\\
$\o_{ik}$ & Whether line $i$ get overloaded when line $k$ trips.\\

% %$tl_i$ & Whether line $i$ exists in the true (real) topology. \\
% $u_i$ & Whether line $i$ exists in the true (real) topology. \\
% %$fl_i$ & Whether line $i$ is fixed in the topology. \\
% $v_i$ & Whether line $i$ is fixed in the topology. \\
% %$sl_i$ & Whether the status information regarding line $i$ is secured. \\
% $w_i$ & Whether the status information regarding line $i$ is secured. \\
% %$el_i$ & Whether line $i$ is excluded from the topology by an exclusion attack. \\
% $p_i$ & Whether line $i$ is excluded from the topology by an exclusion attack. \\
% %$il_i$ & Whether line $i$ is included in the topology by an inclusion attack. \\
% $q_i$ & Whether line $i$ is included in the topology by an inclusion attack. \\
% %$ml_i$ & Whether line $i$ is considered (though it may not exist) in the topology. \\
% $k_i$ & Whether line $i$ is considered (though it may not exist) in the topology. \\
\hline
\end{tabular}
%\normalsize
\vspace{-6pt}
\end{table}

\vspace{3pt}
\noindent\textbf{Pre-attack Case with the SCOPF Model:}
%%%%%%%%--------------------------------------------------------------%%%%%%%%%%%%

To model power flow analysis, we used DC power flow model. % which is discussed in section~\ref{DCPowerFlowModel}. 
According to the discussion, transmission lines are considered lossless thus the admittance is calculated only from the reactance. We denote the power flow through line $i$ using $ P^L_i $ and the admittance of the line using $d_i$. The buses connected through the line $i$ where  are indicated by $f_i$ (\emph{from-bus}) and $e_i$ (\emph{to-bus}) where $1 \le i \le l$, $1 \le f_i, e_i \le b$, and $b$ represents the number of buses. The basic power flow equation is shown in Equation (\ref{pre_btheta}).  Here $\theta_j$ represents the state of bus j which is the phase angle. Therefore, $\theta_{f_i}$ and $\theta_{e_i}$ denote the phase angles of the buses $f_i$ and $e_i$, respectively. The equation specifies that power flow $P^L_i$ is dependent on the difference between the connected buses' phase angles and line admittances.  As in the DC model, each state corresponds to a bus, so the number of states $n$ is equal to $b$. 

Let $P^B_j$ be the power consumption of bus $j$ which is the summation of the power flows of the lines connected to this bus. Let ${\mathbb L}_{j,\mathit{in}}$ and ${\mathbb L}_{j,\mathit{out}}$ be the sets of incoming and outgoing lines of bus $j$, respectively. The Equation (\ref{pre_sumpline}) shows the power consumption at bus $j$. The power consumption at a bus is also equal to the load at the bus minus the power injected to the bus by the generators connected to this bus. The Equation (\ref{pre_buspower}) shows the power consumption,  where $P^G_j$ and $P^D_j$ denote the generated power and load power of bus $j$, respectively. If no generator is connected to bus $j$, then $P^G_{j} = 0$. Similarly, if no load is presented at bus $j$, then $P^D_{j} = 0$. The load of the bus $j$ should be within the rated capacity. Considering $P^{D}_{j,min}$ and $P^{D}_{j,max}$ as the minimum and maximum limit for the load at bus j,  Equation (\ref{pre_loadlimit}) shows the constraint for the limit. Besides, the generators also have to maintain the rated capacity. As for the Equation (\ref{pre_genlimit}), given $P^{G}_{j,min}$ and $P^{G}_{j,max}$ as the minimum and maximum generation capacity of the generators, the output should be within these operating limits while running. 

%%%%%%---------------------------------------------------------------------------------%%%%
\begin{table}[t]
\vspace{-12pt}
\caption{Formalization of Pre-Attack Power Balances} 
\label{Tab:SCOPF_preattack} 
\vspace{-6pt}
\centering
\begin{tabular}{p{\columnwidth}}
\hline
%%%%%%%%%%%%%%%%%%%%%%%%%%%%%%%%%%%%%%%
\begin{flushleft}Line Power Flow:\end{flushleft}
\begin{equation}
\label{pre_btheta}
\forall_{1 \le i \le l}~~ P^L_i = d_i (\theta_{f_i} - \theta_{e_i})
\end{equation}
%%%%%%%%%%%%%%%%%%%%%
\begin{flushleft}Bus Power Consumption:\end{flushleft}
\begin{equation}
\label{pre_sumpline}
\forall_{1 \le j \le b}~ { {P}^B_{j}} = {\sum_{i \in {\mathbb L}_{j, \mathit{in}}} { P}^L_{i}} - {\sum_{i \in {\mathbb L}_{j, \mathit{out}}} { P}^L_{i}}
\end{equation}
\begin{equation}
\label{pre_buspower}
\forall_{1 \le j \le b}~~ P^B_j = P^D_{j} ~- P^G_{j}
\end{equation}
%%%%%%%%%%%%%%%%%%
\begin{flushleft}Load and Generation Limit:\end{flushleft}
\begin{equation}
\label{pre_loadlimit}
\forall_{1 \le j \le b}~ { {P}^{}_{j, min}} \le { {P}^D_{j}} \le{ {P}^{D}_{j, max}} 
\end{equation}
\begin{equation}
\label{pre_genlimit}
\forall_{1 \le j \le b}~ ({ {P}^{G}_{j, min}} \le { {P}^G_{j}} \le{ {P}^{G}_{j,max}})  \vee  ({ {P}^G_{j}}=0)
\end{equation}
%%%%%%%%%%%%%%%%%%
\begin{flushleft}Contingency Line Flow using LODF\end{flushleft}
\begin{equation}%\nonumber
\label{pre_lodf}
\forall_{1 \le i\le l}\forall_{1 \le k\le l}~{{P}}^{L}_{(i,~k)}~=~P^L_i+P^L_k \times LODF_i^k 
\end{equation}
%%%%%%%%%%%%%
%%%%%%%%%%%%%%%%%%
\begin{flushleft}Line Power Limit (\textit{Pre-contingency \& Post-contingency Periods}):\end{flushleft}
\begin{equation}%\nonumber
\label{pre_linelimit}
\forall_{1 \le i \le l}~~ | {P}^L_i |~~\le P^L_{i, \mathit{max}}
\end{equation}
\begin{equation}%\nonumber
\label{pre_linconlimit}
\forall_{1 \le i\le l}\forall_{1 \le k\le l}~~ | {{P}}^{L}_{(i,~k)} |~~\le P^L_{i, \mathit{max}}
\end{equation}
%%%%%%%%%%%%%
\begin{flushleft}SCOPF Generation Cost:\end{flushleft}
\begin{equation}%\nonumber
\label{pre_cost}
T_\mathit{SCOPF}= \sum_{1\le j \le b} \mathcal{C}(P_j^G) 
\end{equation}
\\\hline
\end{tabular}
\normalsize
\end{table}
%%%%%%%%--------------------------------------------------------------------%%%%%%%%%%%%%

If line $k$ trips, the post fault power flow through line $i$  is represented as $P_{(i,k)}^L$ which calculated using Equation (\ref{pre_lodf}). Here, $LODF^k_i$ represents the line outage distribution faction (LODF) of line $i$ when line $k$ trips. Interested readers can find a brief discussion about LODF calculation in Appendix~\ref{Appendix:LODF}. Each transmission line has a limited power flow capacity. Assume $P^L_{i, \mathit{max}}$ as the line upower flow capacity. Therefore Equation (\ref{pre_linelimit}) shows the line power flow capacity limit for each line. As the system is running on SCOPF, all the contingency conditions are considered during the optimization process.  Thus, as per Equation (\ref{pre_linconlimit}) when the line $k$ trips, power flow through any line $i$ should be within the rated capacity.

Let $\mathcal{C}_j(.)$ denote the cost function of the generator connected to bus $j$, which takes the total power generated as the input and returns the cost. Usually, $\mathcal{C}_j(.)$ is a strictly increasing convex function. 
%%Many electric utilities prefer to represent their generator cost functions as piece-wise linear equations, \ie single or multiple segment linear cost functions~\cite{Wood96}.%%
The cost function in this paper is considered as $\mathcal{C}_j(\hat{P}^G_j) = \alpha + \beta \hat{P}^G_j$, where $\alpha$ and $\beta$ represent the cost-coefficients for a specific generator. 
% As previousely mentioned, in SCOPF, the objective is to minimize the total generation cost based on expected or estimated loads at different buses. Satisfying all other constraints,
SCOPF finds the optimum set of a generation with the minimum generation cost, $T_\mathit{SCOPF}$ as shown in Equation(\ref{pre_cost}).
%We use the notation $\mathit{SCOPF}$ to denote the conjunction of the SCOPF constraints, as we have described above, which we will use below to formalize the impact of stealthy attacks on SCOPF.

\vspace{3pt}
\noindent\textbf{Attack Attributes | CPS Attack Mapping:}

In the DC model, two measurements, forward and backward line power flows, can be measured for each line. Also, power consumption can be measured for each bus. Therefore, for a power system with $l$ number of lines and $b$ number of buses, there is $2l + b$ (\ie $m = 2l + b$) number of potential measurements at  most. 
%Though a significantly smaller number of measurements are sufficient for state estimation, redundancy is provided to identify and filter bad data. 
%We define $t_i$ to denote whether potential measurement~$i$ ($1 \le i \le m$) is taken. 
Note that forward and backward power flows of the line $i$ are denoted with the index $i$ and ${l + i}$ respectively, while power consumption of bus $j$ is shown with the index $2l + j$.

%%In false data injection attacks, one or more states of the system can be infected. We define $c_j$ to denote whether state~$j$ is infected (\ie changed to an incorrect value). Parameter $a_i$ denotes whether measurement $i$ (${1 \le i \le m}$) is required to be altered (by injecting false data) for the attack. If any measurement at bus $j$ is required to be changed, $b_j$ becomes true. 

%The attacker may not be able to alter a measurement due to inaccessibility or existing security measures.  We define $r_i$ to denote whether measurement $i$ is accessible to the attacker. We also define $s_i$ to denote whether the measurement is secured (\ie data integrity is protected) or not.  An attacker often needs to know the admittance of the necessary transmission lines to inject the false data at the right amount, so that the attacker can remain undetected. We assume that the attacker has complete knowledge about the admittance o the transmission lines.%

\vspace{3pt}
\noindent\textbf{False Data Injection to Measurements:}

Assume $\Delta P^L_i$ as the change in line $i$ power flow measurement.  According to the Equation (\ref{pre_sumpline}), $\Delta P_j^B$ at bus $j$ depends on the changes made in power flow measurements of the lines connected to the bus $j$. ${\Delta P}^L_{i} \neq 0$ specifies that if measurements $i$ and ${l + i}$ corresponding to line $i$ are taken (\ie $t_i$ and $t_{l + i}$), they are required to be changed by ${\Delta P}^L_{i}$.  Likewise, the power consumption measurement at bus $j$ is required to be changed when ${\Delta P}^B_{j} \neq 0$ and it is taken. These conditions can be formalized as Equations (\ref{Eq_M_FDI_17}). Conversely, measurement $i$ is altered, if it is taken and the corresponding power measurement is changed. This constraint can be formalized as Equation (\ref{Eq_M_FDI}).
%

%%%%%%------------------------------------------------------------------------------------------%%%%%%%%%
\begin{table}[t]
\caption{Formalization of Attack Attributes} 
\vspace{-6pt}
\label{Tab:cps_attack} 
\centering
\begin{tabular}{p{\columnwidth}}
\hline
%%%%%%%%%%%%%%%%%%%%%%%%%%%%%%%%%%%%%%%
\begin{flushleft}Line Power Flow:\end{flushleft}
\begin{equation}
\label{Eq_M_FDI}
\begin{aligned}
&\forall_{1 \le i \le l}~ ({{\Delta P}^L_{i}}   \neq 0) \rightarrow (t_i \rightarrow a_i) \wedge (t_{l + i} \rightarrow a_{l + i}) \\
&\forall_{1 \le j \le b}~~ ({{\Delta P}^B_{j}}   \neq 0) \rightarrow (t_{2 l + j} \rightarrow a_{2 l + j})
\end{aligned}
\end{equation}
%%%%%%%%%%%%%%%%%%%%%
\begin{flushleft}Attack Plan Properties:\end{flushleft}
\vspace{-3pt}
\begin{equation}
\label{Eq_M_FDI_17}
\begin{aligned}
&\forall_{1 \le i \le l}~~ a_i \rightarrow t_i \wedge ({{\Delta P}^L_{i}} \neq 0) \\
&\forall_{1 \le i \le l}~~ a_{l + i} \rightarrow t_{l + i} \wedge ({{\Delta P}^L_{i}} \neq 0) \\
&\forall_{1 \le j \le b}~~ a_{2l + j} \rightarrow t_{2l + j} \wedge ({{\Delta P}^B_{j}} \neq 0)\\
&\forall_{1 \le j \le n}~~ c_j \rightarrow ({\Delta \theta}_j \neq 0)\\
\end{aligned}
\end{equation}\begin{flushleft}Attack Access Capability:\end{flushleft}
\begin{equation}
\label{securedandaccesss}
\forall_{1 \le i \le m}~~ a_i \rightarrow r_i \wedge \neg s_i
\end{equation}
%%%%%%%%%%%%%%%%%%
\begin{flushleft}Attack Resources\end{flushleft}
\vspace{-3pt}
\begin{equation}
\label{hlimits}
\begin{aligned}
& \forall_{1 \le i \le l}~~ a_i \rightarrow h_{f_i} \\
& \forall_{1 \le i \le l}~~  a_{l + i} \rightarrow h_{e_i} \\
& \forall_{1 \le j \le b}~~ a_{2l + j} \rightarrow h_j
\end{aligned}
\end{equation}
\begin{equation}
%\left(\sum_{b_j} 1\right) \le T_{CB}
\label{TBlimit}
\sum_{1 \le j \le b} {h_j} ~\le T_{B}
\end{equation}
%%%%%%%%%%%%%%%%
\\\hline
\end{tabular}
\normalsize
\vspace{-6pt}
\end{table}

%%%%%%%%----------------------------------------%%%%%%%%%%%%%
%\noindent\textbf{Limited Capabilities.}
Attacker's physical or remote access to the measurements are of the key factors to false data injection to a measurement. However, if a measurement is secured (\ie data integrity protected) the FDI attack will not be successful even if the attacker has access to the measurement. Hence, the attacker will only be able to change the measurement $i$ if Equation (\ref{securedandaccesss}) becomes true.
Due to limitation, an attacker is able to attack only some of the substations at a particular time.  In case a measurement of a substation needs to be changed that substation must be either compromised or accessed.
This constraint is represented in Equation (\ref{hlimits}).  %
Let $T_{B}$ be the maximum number of compromisable substations, then the number of compromised substation should be within $T_{B}$ as shown in Equation (\ref{TBlimit}) .
%
%%%%%%%%%%%%%%%%%%%%%%%%%%%%%%%%%%%%%%

\vspace{3pt}
\noindent\textbf{Impact of FDI Attack on the SCOPF:}

%%%%%%%%%%%%%
\begin{table}[t]
\caption{Formalization of FDI Attack on Physical System} 
\vspace{-6pt}
\label{Tab:SCOPF_preattack} 
\centering
\begin{tabular}{p{\columnwidth}}
\hline
%%%%%%%%%%%%%%%%%%%%%%%%%%%%%%%%%%%%%%%
\begin{flushleft}Attacked Line Power Flow:\end{flushleft}
\begin{equation}
\label{att_btheta}
\forall_{1 \le i \le l}~~{\Delta P}^L_i = d_i ({\Delta \theta}_{f_i} - {\Delta \theta}_{e_i})
\end{equation}
\begin{equation}
\label{att_deltaline}
 \forall_{1 \le i \le l}~~{\bar P}^L_i =  P^L_i + \Delta P^L_i 
\end{equation}
%%%%%%%%%%%%%%%%%%%%%

%%%%%%%%%%%%%%%%%%
\begin{flushleft}Attacked Bus and Load Data:\end{flushleft}
\begin{equation}
\label{att_delbuspower}
\forall_{1 \le j \le b}~ {\Delta {P}^B_{j}} = {\sum_{i \in {\mathbb L}_{j, \mathit{in}}} {\Delta {P}}^L_{i}} - {\sum_{i \in {\mathbb L}_{j, \mathit{out}}} {\Delta {P}}^L_{i}}
\end{equation}
\begin{equation}
\label{att_bceqbload}
\forall_{1 \le j \le b}~ {\Delta {P}^B_{j}} ={\Delta {P}^D_{j}}
\end{equation}
\begin{equation}
\label{att_atload}
\forall_{1 \le j \le b}~ {\bar {P}^D_{j}} = { {P}^D_{j}}+ {\Delta {P}^D_{j}}
\end{equation}
\begin{equation}
\label{att_attloadlimit}
\forall_{1 \le j \le b}~ { {P}^{D,min}_{j}} \le {\bar {P}^D_{j}} \le{ {P}^{D,max}_{j}} 
\end{equation}

%%%%%%%%%%%%%%%%%%
\begin{flushleft}Limit of Load Injection :\end{flushleft}
\begin{equation}
\label{att_perdelta}
\forall_{1 \le j \le b}~ |~{\Delta {P}^D_{j}}~|  \le  P^D_j \times \delta_b 
\end{equation}
%%%%%%%%%%%%%%%%%%%%%%%%%%%%%%%%%%%%%%%%%%%
\\\hline
\end{tabular}
\normalsize
\vspace{-6pt}
\end{table}
%%%%%%%%--------------------------------------------------------------------%%%%%%%%%%%%%

From Equation (\ref{pre_btheta}) it is clear that, a change of $P^L_i$ needs to change based on the changes in the state ${f_i}$ ($\theta_{f_i}$) and/or state ${e_i}$ ($\theta_{e_i}$) or vice versa. Thus, ${\Delta P}^L_i$ is calculated from Equation (\ref{att_btheta}). If ${\Delta \theta}_{f_i} \neq 0$ (or ${\Delta \theta}_{e_i} \neq 0$), then it is obvious that state ${f_i}$ (or ${e_i}$) is changed (\ie infected). However, if both of the states change equally, ${\Delta P}^L_i$ still remains zero. As shown in the Equation (\ref{att_deltaline}) the attacked line power $\bar P_i^L$ is the summation of original line power $ P_i^L$ and the injected amount $ {\Delta P}^L_i $. Any variable with the notation  ( $\bar {           } $ ) over it indicates that, the amount is injected through the FDI attack, which may not be equal to the actual (field) value. 

The change in the bus consumption $\Delta P^B_j$ due to attack is represented by the Equation (\ref{att_delbuspower}) which is the sum of changes in the power flow of connected lines. According to Equation~(\ref{pre_buspower}), ${\Delta P^B_{j}} \neq 0$ means that there is a power generation and/or load change at the bus. In this paper, it is assumed that any change in a bus power consumption measurement can only be made though a change in the load; in other words ${\Delta P}^G_{j} = 0$. Therefore, changes in the bus power consumption specifies the load changes in those buses as shown in Equation (\ref{att_bceqbload}). This is due to the fact that, generator power measurements are pretty much well-defined and are changeable only by the grid operators. Usually, after performing the state estimation, if there are any load changes, SCOPF is performed and the result would determine whether (and which) changes in the generation are necessary for the optimal operation. As shown in the Equation (\ref{att_atload}), the attacked bus load $\bar P_j^D$ is the summation of original bus load $ P_j^D$ and the injected false load, $ {\Delta P}^D_j $. 

Though properly injected any large abrupt change in the measurement will bypass the bad data detector, however, may create suspicion to the system operator. Thus to make the attack stealthy enough, the change in the bus load is limited to a certain percent of the original load. The Equation (\ref{att_perdelta}) shows that the changed bus load $\Delta P^D_j$ is within the $\delta_b$ percent of the pre-attack value.  Besides, the total attacked load is within the rated limit as shown in Equation (\ref{att_attloadlimit})

\vspace{3pt}
\noindent\textbf{Running SCOPF on Corrupted Loads}
%%%%%%---------------------------------------------------------------------------------%%%%

%\noindent\textbf{Security Constrained Optimal Power Flow (SCOPF):}\vspace{-3pt}
%\label{SubSec:Model_OPF}
\vspace{3pt}
%%%%%%%%%%%%%%%%%%%%%%%%%%%%%%%%%%%%%%%%%%%%%%%%%%%%%%%%%%%%%%%%%%%%%%%

%The objective of the SCOPF is to optimally control the generation according to the load requirement and contingency conditions.
Due to the state estimation of attacked load data, consider $\hat{P}^G_j$ as the altered power generation  by the generator connected at $j$ which follows the equality of the generation and expected load as shown in Equation (\ref{ac_totalsum}). Here any character with the notation ( $\hat{    }$ ) represents the actual filed data after the attack. Again, the generation $\hat{P}^G_{j}$ is withing the rated generation limit as per the Equation (\ref{ac_genlimit}).

%%%%%%%%%%%%%%%%%%%%
\begin{table}[t]
\caption{Formalization of Generation Dispatches on Corrupted Load} 
\label{Tab:SCOPF_attack} 
\vspace{-6pt}
\centering
\begin{tabular}{p{\columnwidth}}
\hline
%%%%%%%%%%%%%%%%%%%%%%%%%%%%%%%%%%%%%%%
\begin{flushleft}Updated Generation:\end{flushleft}
\vspace{-3pt}
\begin{equation}%\nonumber
\label{ac_totalsum}
\sum_{1 \le j \le b} \hat{P}^G_j = \sum_{1 \le j \le b} \bar{P}^D_j
\end{equation}
%%%%%%%%%%%%%%%%%%%%%
\begin{flushleft}Generation Limit:\end{flushleft}
\begin{equation}
\label{ac_genlimit}
\forall_{1 \le j \le b}~ ({ {P}^{G}_{j, min}} \le {\hat{{P}}^G_{j}} \le{ {{P}}^{G}_{j,max}})  \vee  ({ \hat{P}^G_{j}}=0)
\end{equation}
%%%%%%%%%%%%%%%%%%
\begin{flushleft}Power Flow Constraints:\end{flushleft}
\begin{equation}%\nonumber
\begin{aligned}
\label{att_buspower}
& \forall_{1 \le j \le b}~~ \bar{P}^B_j = \bar{P}^D_{j} ~- \hat{P}^G_{j}\\
& \forall_{1 \le j \le b}~~ \bar{P}^B_j = \sum_{i \in {\mathbb L}_{j, \mathit{in}}} \bar{P}^L_{i} ~- \sum_{i \in {\mathbb L}_{j, \mathit{out}}} \bar{P}^L_{i} \\
&\forall_{1 \le i \le l}~~{\Delta \bar P}^L_i = d_i ({\Delta \bar \theta}_{f_i} - {\Delta \bar \theta}_{e_i})
\end{aligned}
\end{equation}
%%%%%%%%%%%%%%%%%%

\begin{flushleft}Expected Line Power Flow (\textit{Pre-contingency \& Post-contingency Periods}):\end{flushleft}
\begin{equation}%\nonumber
\begin{aligned}
\label{att_linelim}
&\forall_{1 \le i\le l}\forall_{1 \le k\le l}~{{\bar P}}^{L}_{(i,~k)}~=~\bar P^L_i+\bar P^L_k \times LODF_i^k \\
&\forall_{1 \le i \le l}~~ \bar{P}^L_i \le P^L_{i, \mathit{max}}\\
&\forall_{1 \le i \le l} \forall_{1 \le k \le l}~~ \bar{P}^L_{i,k} \le P^L_{i, \mathit{max}}
\end{aligned}
\end{equation}
%%%%%%%%%%%%%
\begin{flushleft}SCOPF Generation Cost Limit:\end{flushleft}
\vspace{-3pt}
\begin{equation}%\nonumber
\label{att_cost}
\sum_{1 \le j \le b} \mathcal{C}_j({\hat{P}^G_j}) ~\le T_\mathit{SCOPF}
\end{equation}
%%%%%%%%%%%%%%%%
\\\hline
\end{tabular}
\normalsize
\vspace{-6pt}
\end{table}
%%%%%%%%--------------------------------------------------------------------%%%%%%%%%%%%%

 Considering this scenario, let $\bar{\theta}$, $ and $ $\bar{P}^B_j$  be the expected phase angle and the power consumption at bus $j$, respectively. All of the power-flow equations in (\ref{att_buspower}) are considered as constraints of SCOPF with expected attacked load and new generation. 
 As the system is running on SCOPF, all the expected line power flows during both pre-contingency and post-contingency must be within the threshold as shown in Equation (\ref{att_linelim}). Here $\bar{P}^L_{i,k}$  denotes the power flow through line $i$ when like $k$  trips which is calculated using the LODF matrix. 
 
 The attacker's target is not to increase the generation cost. Thus the total cost should be less or equal to the pre-attack generation cost. Thus Equation (\ref{att_cost}) shows the cost constraint with the new set of generations. 

\vspace{3pt}
\noindent\textbf{Sub-optimal Generation Dispatches for Actual Loads and Ultimate Attack Goal:}

%\noindent\textbf{Running SCOPF on Actual Loads}\vspace{-3pt}
The generations are updated by EMS to optimize the system which is done based on the false load data. However, the actual loads are still the same as the pre-attack condition which is $P^D_j$. Thus, the system is now running with the inaccurate generation $\hat P^G_j$ which is not optimal anymore. After the attack, considering $\hat P^B_j$ be the actual bus consumption, the SCOPF constraints with actual data are shown in Equation (\ref{ac_scopf}).

%%%%%%---------------------------------------------------------------------------------%%%%
\begin{table}[t]
\caption{Formalization of Generation Dispatches on Actual Loads} 
\label{Tab:SCOPF_actual} 
\vspace{-6pt}
\centering
\begin{tabular}{p{\columnwidth}}
\hline
%%%%%%%%%%%%%%%%%%%%%%%%%%%%%%%%%%%%%%%
%%%%%%%%%%%%%%%%%%%%%
\begin{flushleft}Bus Power  Consumption:\end{flushleft}
\begin{equation}
\begin{aligned}
\label{ac_scopf}
&\forall_{1 \le j \le b}~~ \hat{P}^B_j = {P}^D_{j} ~- \hat {P}^G_{j}\\
&\forall_{1 \le j \le b}~ { \hat{P}^B_{j}} = {\sum_{i \in {\mathbb L}_{j, \mathit{in}}} { \hat{P}}^L_{i}} - {\sum_{i \in {\mathbb L}_{j, \mathit{out}}} { \hat{P}}^L_{i}}
\end{aligned}
\end{equation}

\begin{flushleft} Actual Line Power Flow: \end{flushleft}
\begin{equation}
\label{lineflow_actual}
\forall_{1 \le i \le l}~~\hat{ P}^L_i = d_i ({ \hat \theta}_{f_i} - \hat{ \theta}_{e_i})
\end{equation}

\begin{flushleft}Attacker's Goal on Line Power Flow (\textit{Pre-contingency \& Post-contingency Periods}):\end{flushleft}
\begin{equation}
\label{ac_normallimit}
\forall_{1 \le i \le l}~~ |{ \hat {P}}^L_i |~~\le P^L_{i, \mathit{max}}
\end{equation}
\begin{equation}%\nonumber
\label{ac_maintarget}
\begin{aligned}
&\forall_{1 \le i\le l}\forall_{1 \le k\le l}~{{\hat P}}^{L}_{(i,~k)}~=~\hat P^L_i+\hat P^L_k \times LODF_i^k \\
&\forall_{1 \le i \le l} ~~  \o_{ik}~~ \rightarrow  ~~~| {{\hat{P}}}^{L}_{(i,~j)} |~~ >  P^L_{i, \mathit{max}} \times \delta_l \\
&\sum_{1 \le i \le l}\sum_{1 \le k \le l} {\o_{i,k}} ~\ge T_{L}\\
\end{aligned}
\end{equation}
%%%%%%%%%%%%%%%%%%%%%%%%%%%%%%%%%%%%%%%%%%%
\\\hline
\end{tabular}
\normalsize
\vspace{-6pt}
\end{table}
%%%%%%%%--------------------------------------------------------------------%%%%%%%%%%%%%

Now the attacker's goal is not to overload any line during the normal operation as shown in Equation (\ref{ac_normallimit}). However, the attacker's ultimate target is to overload the lines during any contingency condition. Thus, if line $k$ trips, few lines will be overloaded by $\delta_l$ percentage among the rest of the running lines. The attacker's goal is to overload a minimum number of $T_L$ lines when any contingency happens. This target is formalized in Equation (\ref{ac_maintarget}).
%%%%%%%%%%%%

\subsection{Implementation}
\label{SubSec:Implem}

The entire constraints as well as the system configuration are encode into SMT~\cite{Moura09}. \emph{Z3~.Net API}~\cite{Z3} is used for encoding the formalization of the proposed FDI model. %, that we have already discussed in Section~\ref{Model}.
The formalizations is mainly encoded using Boolean (\ie for logical constraints) and Real (\eg for the relation between power flows or consumption with states) terms. A text file (\emph{input} file) is used to import the system configurations and constraints.

The verification result of the model execution in Z3 is as either satisfiable (\emph{sat}) or unsatisfiable (\emph{unsat}). The \emph{unsat} result means that there is no attack vector for the problem that satisfies the constraints. If the result is \emph{sat}, the attack vector is extracted from the assignments of the variables, $a_{i}$s (and/or $h_i$s), which show the set of measurements to be altered (or the set of buses to be compromised). 
%The final results  of the model are also saved in a text file (\emph{output} file).

%%%%%%%%%%%%%%%%%%
\subsection{Example Case Studies}
\label{SubSec:Model_Examp}

\begin{figure}[t]
\vspace{-6pt}
\begin{center}
\includegraphics[width=\columnwidth]{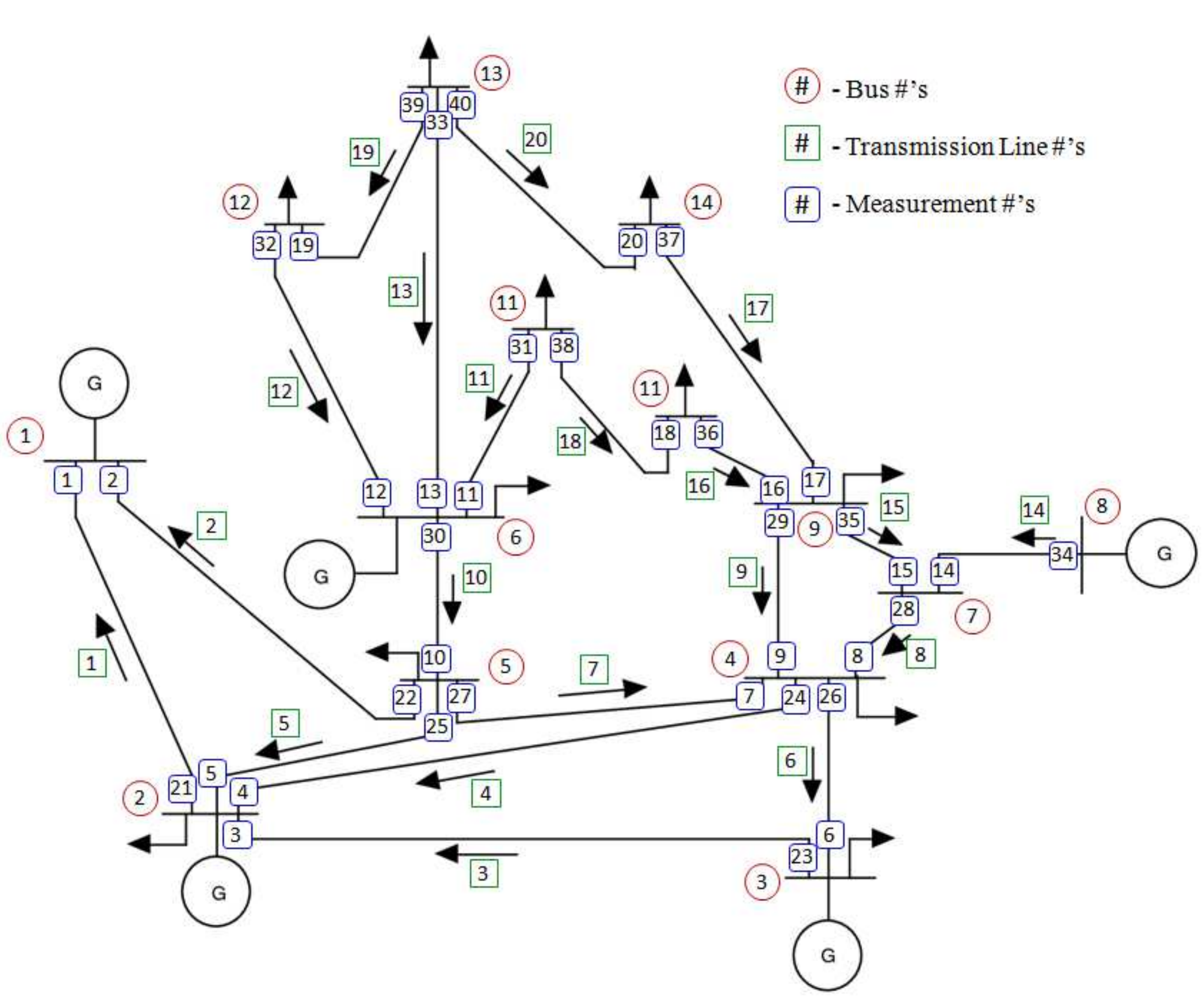}
%\vspace{-9pt}
\caption{IEEE 14-bus test system topology~\cite{rahman2018impact}. The buses, transmission lines, and measurements are numbered.}
\label{Fig_5_Bus}
\end{center} 
\vspace{-12pt}
\end{figure}

Here, we present a few cases to demonstrate our the execution of our proposed formal model to identify UFDI attacks impacting contingency analysis. In these examples, IEEE 14-bus system (Fig.~\ref{Fig_5_Bus}) is studied.

\vspace{3pt}
\noindent\textbf{Input:}

%The input regarding the line information is shown in Table~\reg{Tab_Input}. %Table~\ref{Tab_Line_Info}. 
%As it can be seen 

The input line information consists of a set of data for each of the lines such as line admittance, line number, end buses (from-bus and to-bus), line capacity (\ie the maximum possible line power flow). According to the input, line 1 connects buses 1 and 2, which has an admittance of 16.90 and the rated capacity is 0.65 per unit (pu) (100 MVA Base). Appendix~\ref{Appendix:CaseStudyInput} (Table~\ref{Tab_Input}) shows partially the input file for the interested readers.  

The input includes necessary information about the buses. The generation boundaries of the generators corresponding to the buses are given. It is assumed that a generation bus has only one generator. The generation power cost function is considered the one shown in Section~\ref{SubSec:Model_SCOPF}. Coefficients $\alpha$ and $\beta$ related to the generators are provided in the input. Note that these coefficient values are considered randomly. The total load of the system is 2.45 pu, (\ie 245 MW, considering a 100 MVA base). 
%%%%%%%%%%%%%%%%%%%%%%%  Shahriar %%%%%%%%%%%%%%%%%%%%%%%%

Since the example bus system has 14 buses and 20 lines, the maximum number of potential measurements is  (14 + 2$\times$20) or 54. Each row of the measurement information includes (i) whether the measurement is taken for state estimation  (ii) whether the measurement is secured and (iii) whether the attacker has the accessibility to alter the measurement. The example input file indicates that the measurement 1 is taken, not secured and can be altered by the attacker. The first 20 measurements represent the forward power flow, the following 20 measurements represent the backward power flow of the 20 lines, and the rest 14 measurements represent the power consumption of the 14 buses. 

The cost constraint in the attack-free condition is \$382, which means according to the current load scenario the SCOPF cost is \$382. The attack vector should find a solution with a generation cost not more than this cost.

 The attacker is allowed to attack up-to 20 measurements those can be distributed among maximum 3 different buses. According to the input, while attacking the measurements the changes in bus consumption should be within the 20\% of the original consumption. Finally, the attacker's target is to find an attack vector which will create 5\% overloading in 5\% transmission lines during the contingency condition.

\vspace{3pt}
\noindent\textbf{Case Study 1:}
%%%%%%%%%%%%%%%%%%%%%%%%%%%%%%%%%%%%%%%%%%%%%%%%%%%%%%%%%%%%%%%%%%%%%%%%%%%%%%%%%%%%%%%%%%%%%%%%%%%%%

In this example, the attacker's objective is to launch a false data injection attack by shifting the loads among the few buses so that system runs at a new operating point where generation cost is not increased but one or more lines will be overloaded if any line trips. Here the attacker has access to all the measurements and none of them are secure means he can inject false data to any of the 54 measurements. The changes in the bus load are limited to 20\% of the original bus load and the target is to overload  5\% of the total lines (i.e., one line) and the overloading amount should be at least 5\%  of the rated capacity. The execution of the model corresponding to this example returns \emph{unsat}, specifying that there is no attack vector which satisfies the attacker's goal by only attacking measurements maximum in 2 buses. The reason behind is that all of the buses except bus 8 are connected to other two or more buses. Thus attacking the state of any of the buses needs to adjust the power flow of the lines connected to that bus. As those lines are connected to at least two different neighboring buses, the measurements in those buses also need to be altered. Besides, bus 8 is the only bus that is connected to only one bus (bus 7). However, as none of the buses have a load, a false data injection attack is not possible. Therefore, a UFDI attack in the measurements distributed in a maximum of two buses is not possible. 

%%%%%%%%%%%%%%%%%%%%%%%%
\begin{figure*}[t]
    \begin{center}
        \subfigure[]{
            \label{Fig_Case_2_attacked}
            \includegraphics[scale=0.1, keepaspectratio=true]{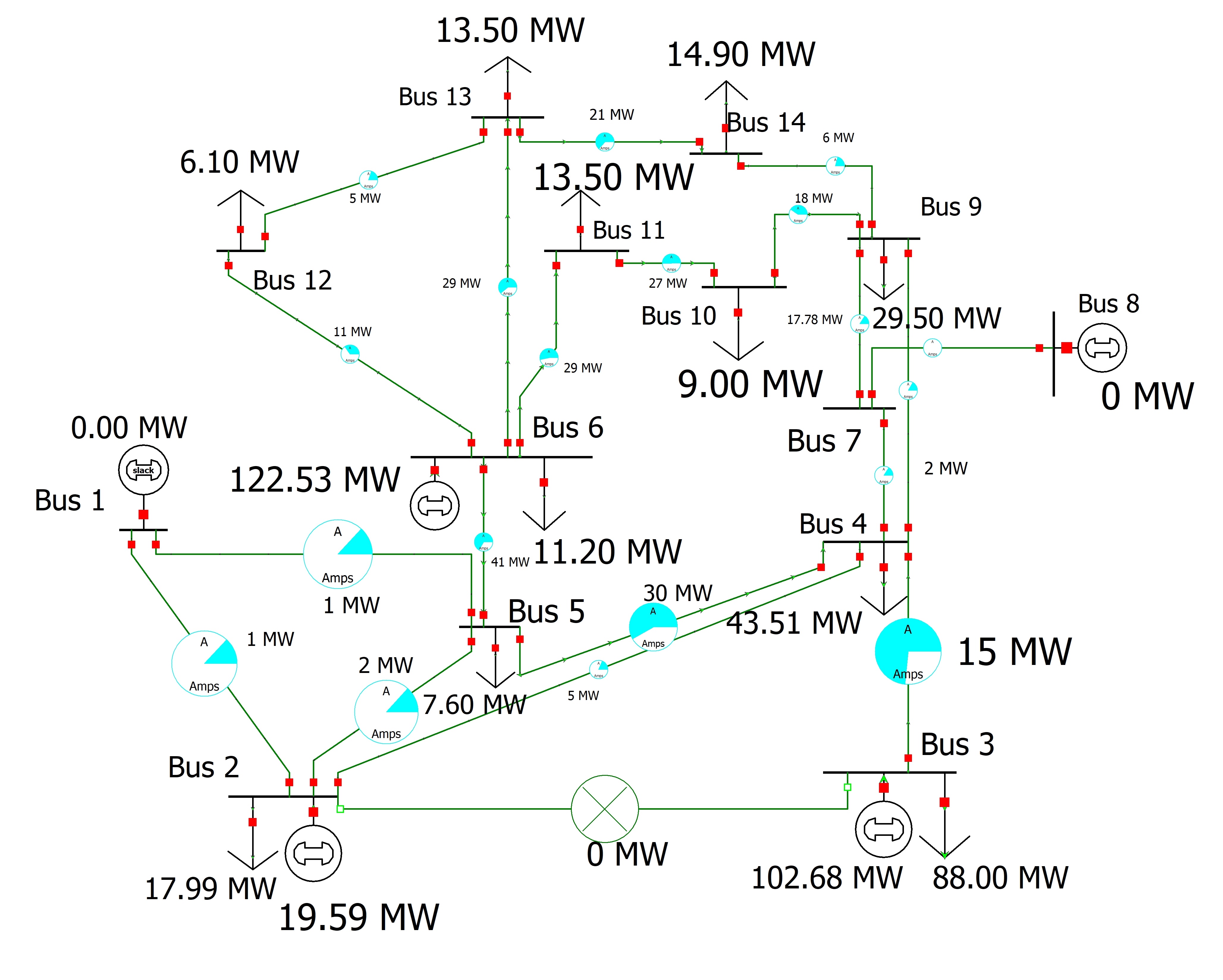}
        }
        \subfigure[]{
            \label{Fig_Case_2_actual}
            \includegraphics[scale=0.09, keepaspectratio=true]{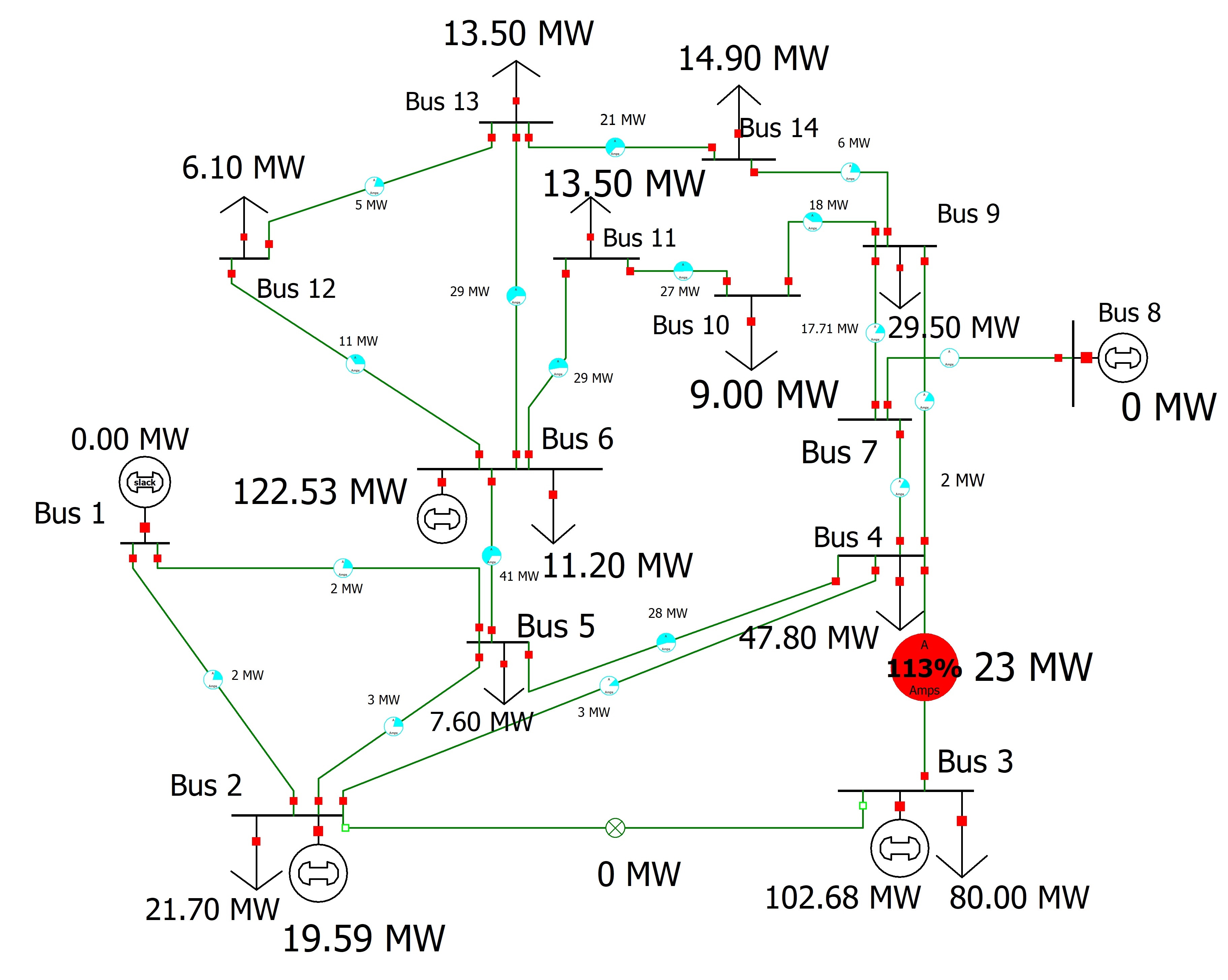}
        }
    \end{center}
    \vspace{-15pt}
    \caption{Power flow of the attack vector for Case Study 2  (a) attacked load: No overloading (b) actual load: line 6 trips, line  3 gets overloaded}
    \label{Graph_Time}
\vspace{-6pt}
\end{figure*}
%%%%%%%%%%%%%%%%%%%%%%%%

%%%%%%%%%%%%%%%%%%%%%%%%
\begin{figure*}[t]
    \begin{center}
        \subfigure[]{
            \label{Fig_Case_3_attacked}
            \includegraphics[scale=0.10, keepaspectratio=true]{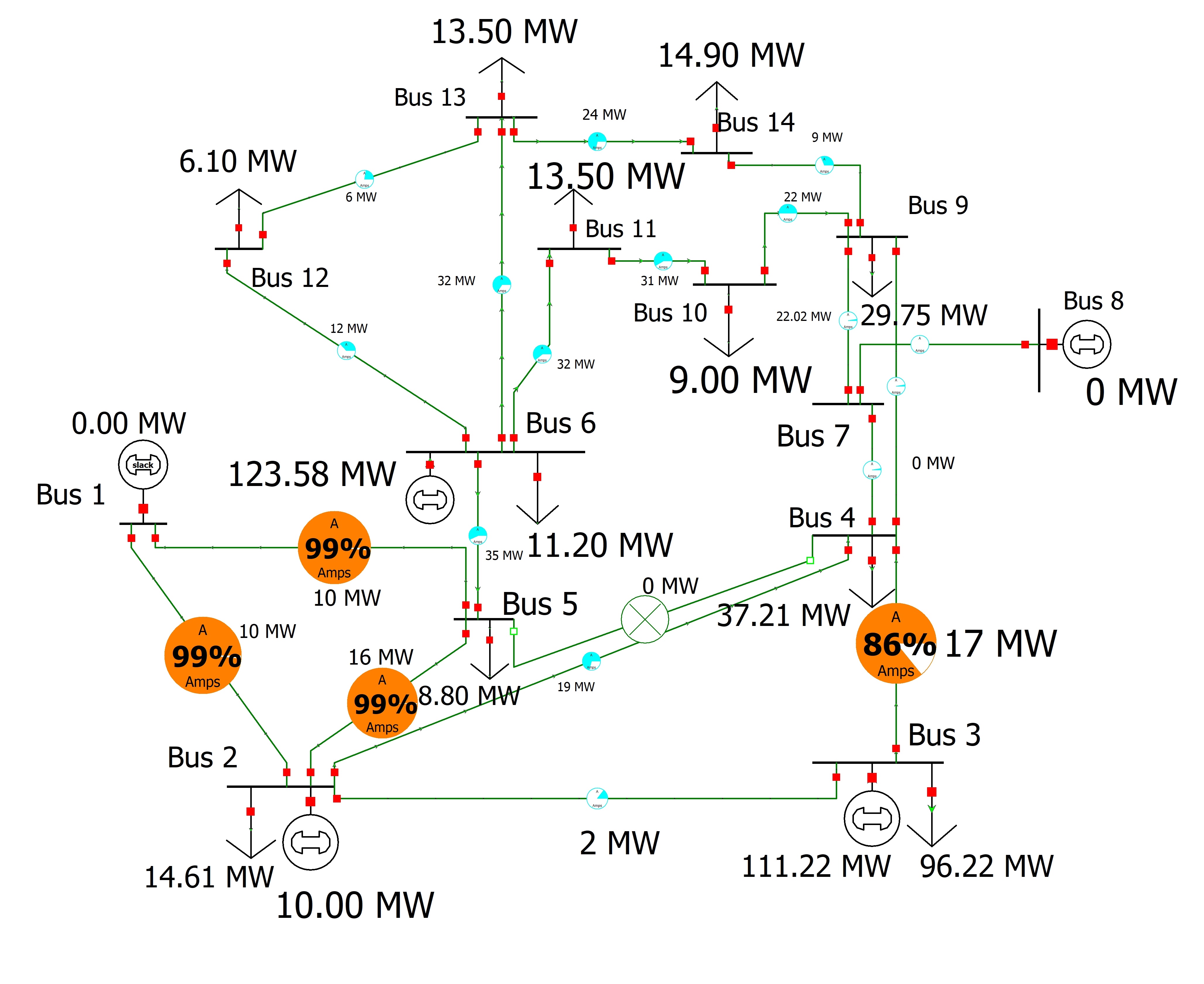}
        }
        \subfigure[]{
            \label{Fig_Case_3_actual}
            \includegraphics[scale=0.10, keepaspectratio=true]{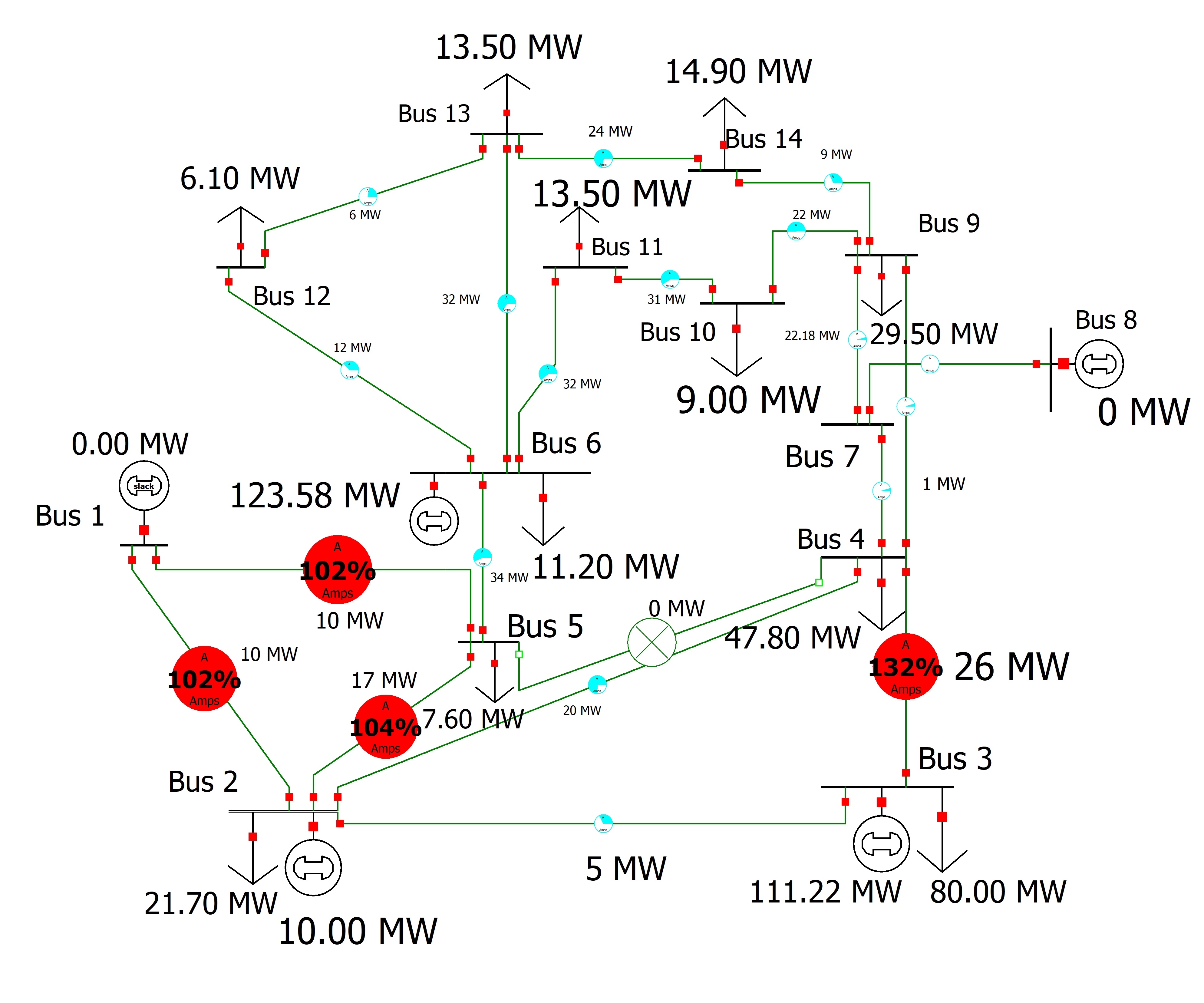}
        }
    \end{center}
    \vspace{-15pt}
    \caption{Power flow of the attack vector for Case Study 3  (a) attacked load: No overloading (b) actual load: line 7 trips, line  1, 2, 5 and 6 get overloaded}
    \label{Graph_Time}
\vspace{-6pt}
\end{figure*}
%%%%%%%%%%%%%%%%%%%%%%%%

%%%%%%%%%%%%%%%%%%%%%%%%%%%%%%%%%%%%%%%%%  Shahriar %%%%%%%%%%%%%%%%%%%%%%%%%%%%%%%%%%%%%%%%%%%%%%%%%

\vspace{3pt}
\noindent\textbf{Case Study 2:}
%%%%%%%%%%%%%%%%%%%%%%%%%%%%%%%%%%%%%%%%%  Shahriar %%%%%%%%%%%%%%%%%%%%%%%%%%%%%%%%%%%%%%%%%%%%%%%%%

In this example, the same attributes used for \emph{case study 1} were kept, except a change in the attacker's capability to attack the measurements. Thus, the attacker has more capability and attacked measurements are distributed in maximum 3 buses instead of 2.  After executing the model  for this example, the results show a \emph{sat} along with returning the assignments to different variables. From these assignments, it is observable that:

\begin{itemize} 
\item Before the attack, the SCOPF solution has a generation cost of \$382 where buses 2,3, and 4 generate 0.284 pu, 0.944 pu, and 1.220 pu, respectively. 
\item To achieve the attacker's goals, it is required to launch a UFDI attack on state 3.
\item For keeping the attack undetected, measurements 3, 6, 23, 26, 42, 43 and 44 requires to be changed. Thus, the attacker needs to inject false on the measurements of line 3 and 6 and buses 2, 3, and 4 given they all are distributed in bus 2, 3, and 4. 
\item From the attack vector, we also see that to attack the state at bus 3, load at bus 3 increases by 0.08 pu and loads at bus 2 and 4 are decreased by 0.037 pu and 0.043 pu respectively. Thus bus load at bus 2, 3 and 4 are changed from 0.8 pu to 0.88 pu, from 0.217 to 0.18 and from 0.478 pu to 0.435 pu, respectively.
\item According to the attacked load scenario, the system calculates the new set of generation dispatches by executing SCOPF. Generators 2, 3, and 6 changes their output from 0.284 pu to 0.196  pu, from 0.944 pu to 1.027 pu, and from 1.220 pu to 1.225 pu, respectively. The new set of dispatch has a generation cost of \$380 which is less than the pre-attack SCOPF generation cost.
\item With false load data and the new set of generations, all the expected line power flows calculated by the EMS are within the rated capacity during the normal condition and the contingency conditions. 
\item However, the loads are actually still the same as the pre-attack condition but the generators have changed output that is not optimal anymore. Thus the real power flow through the lines is different than the expected ones. Although the actual line power flows are within the rated capacity during the normal condition, when line 3 trips, the power flow through line 6 becomes 0.227 pu where the rated capacity of the line is 0.20 pu. The overloading amount is 13.5\%, and thus line 3 will trip as soon as line 6 trips which is unexpected for the EMS. 
\end{itemize}

The PowerWorld~\cite{powerworld} simulator is used to verify the attack vectors from the proposed model. Additionally, the model with the lossless DC power system is simulated in order to verify that there is no overloading with the attacked load and new generation data during both normal and contingency period. The figure \ref{Fig_Case_2_attacked} shows that when line 3 (between buses 2 and 3) trips there is no overloading among any of the lines.  However, Fig.~\ref{Fig_Case_2_actual} shows the the scenario with the new generation and actual load, where we see when line 3 trips, line 6 becomes overloaded by 13.5\%. Conclusively acquiring the same result shown in the model provided.

%Consequently, thanks to this case study, 
In this case study, it is observed that in order to launch a successful UFDI attack on CA, %of the discussed 14 bus power system, 
the attacker needs to attack the measurements that are distributed in at least 3 different buses. Having in mind that the amount of change in the loads should be at least 10\% of the original value.
%%%%%%%%%%%%%%%%%%%%%%%%%%%

%%%%%%%%%%%%%%%%%%%%%%%%
\begin{figure*}[t]
    \begin{center}
    %\hspace{-12pt}            
        \subfigure[]
        {
        \label{Eval_1_1}
            \includegraphics[scale=0.5, keepaspectratio=true]{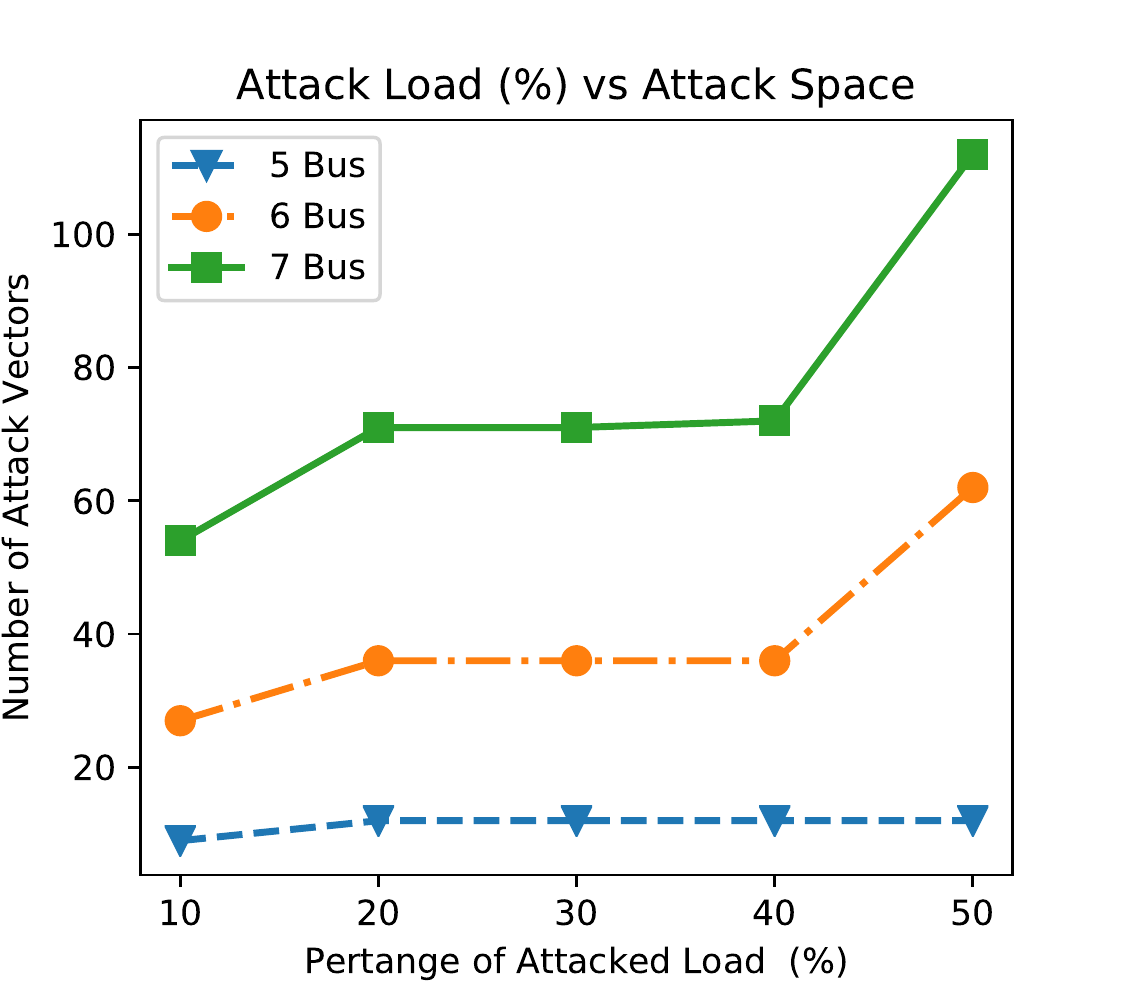}
        }
        \subfigure[]
        {
        \label{Eval_1_2}
            \includegraphics[scale=0.5, keepaspectratio=true]{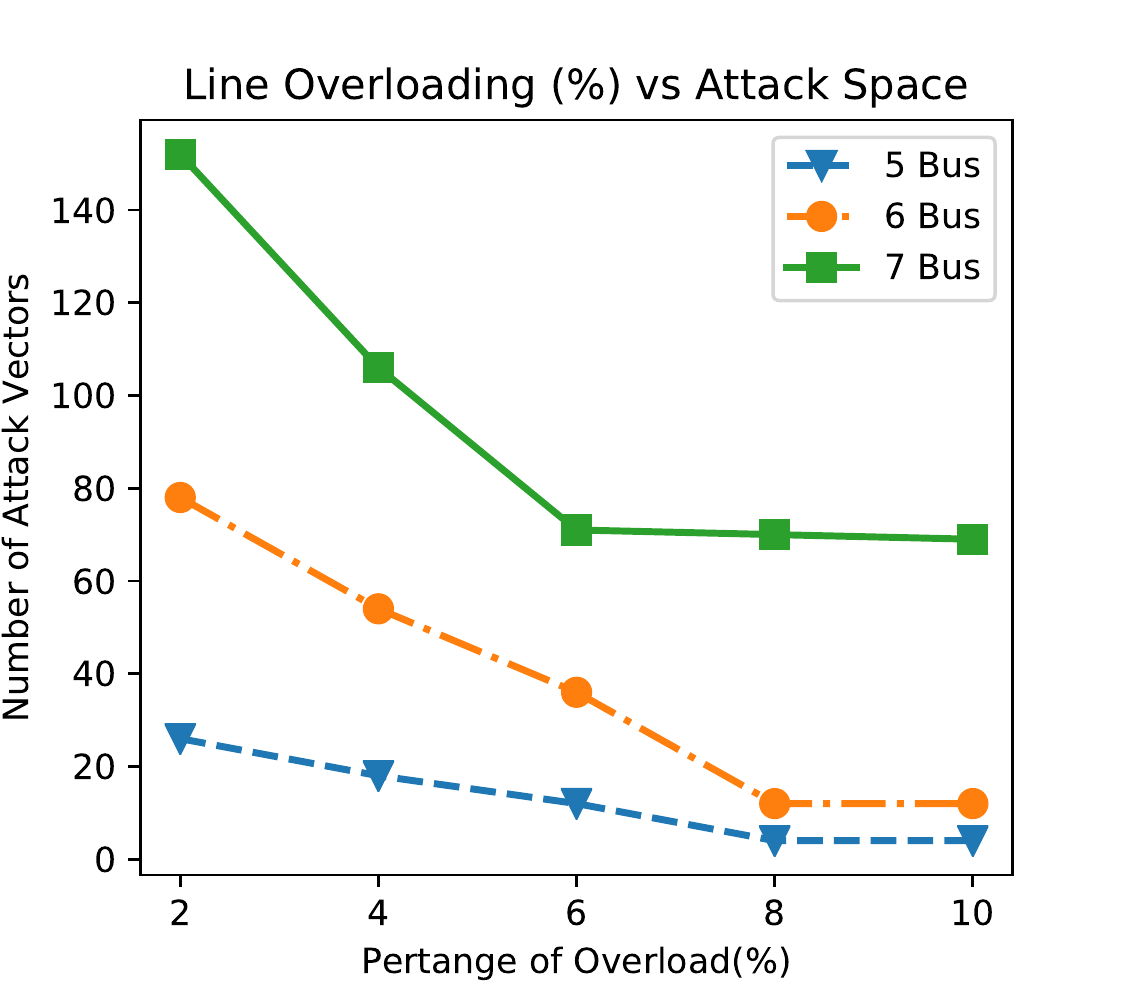}
        }
        \subfigure[]
        {
        \label{Eval_1_3}
            \includegraphics[scale=0.5, keepaspectratio=true]{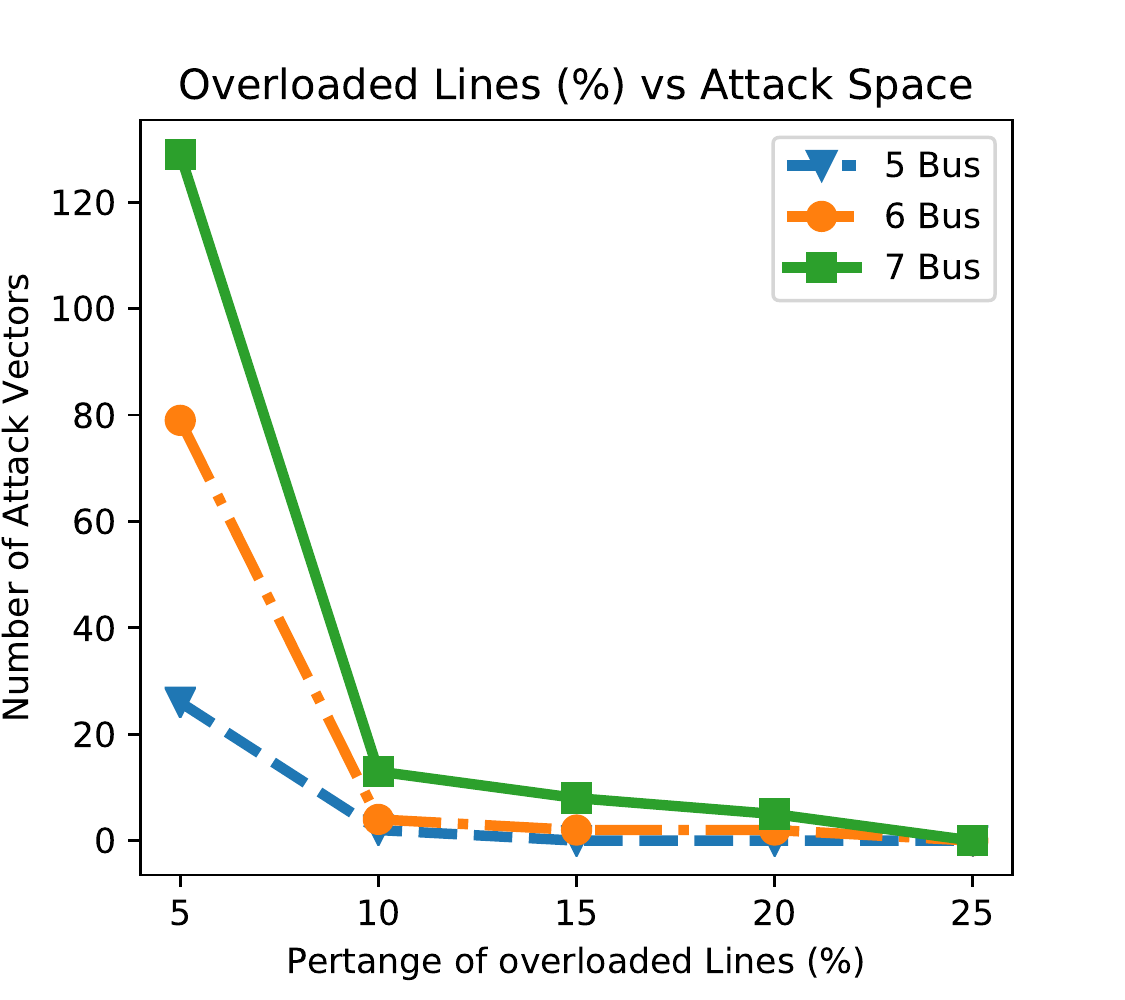}
        }
    \end{center}
    \vspace{-15pt}
    \caption{Evaluation of the model's performance : (a) Impact of attacker's capability (percentage of attacked load), (b) Percentage of line overloading amount, and (c) percentage of overloaded lines}
    \label{Fig:Eval_1}
\vspace{-12pt}
\end{figure*}

\vspace{3pt}
%%%%%%%%%%%%%%%%%%%%%%%%
\noindent\textbf{Case Study 3:}
%%%%%%%%%%%%%%%%%%%%%%%%%%%%%%%%%%%%%%%%%  Shahriar %%%%%%%%%%%%%%%%%%%%%%%%%%%%%%%%%%%%%%%%%%%%%%%%%

In this example, the attacker's objective is to launch a false data injection attack to overload at least 25\% of the total lines (4 lines) within overloading amount of 2\% of the rated capacity at least. The attacker can change the load up to 50\% of the actual value.  The execution of the model corresponding to this example returns \emph{sat} along with the assignments to different variables of the model. From the assignments, we find that:

\begin{itemize} 

\item To achieve the attacker's goals, it is required to execute UFDI attack on states 3, 4, 7, and 8.
\item In order to keep this attack undetected, it is required to alter measurements in buses 2, 3, 4, 5, 7, and 9.  
\item From the attack vector, we see that to attack the states, loads at buses 3, 5, and 9 are increased by 0.162 pu, 0.012, and 0.003 pu, respectively, and loads at bus 2 and 4 are reduced by 0.071 pu and 0.106 pu, respectively. 
\item According to the attacked load scenario, the system calculates the new set of generation running the SCOPF algorithm. Generator 2,3, and 6 change their output from 0.284 pu to 0.1  pu, from 0.944 pu to 1.112 pu, and from 1.220 pu to 1.235 pu, respectively. The new set of dispatch has a generation cost of \$378, which is less than the pre-attack SCOPF generation cost.
\item With the false load data and the new set of generations, all the expected line power flows calculated by EMS are within the rated capacity during the normal condition and the contingency conditions. 
\item However, with the actual loads and new non-optimal generations, the line power flows are within the rated capacity during the normal condition.  However,  when line 7 trips, the power flow through the line 1, 2, 5 and 6  get overloaded by 2\%, 2\%, 3.75\%, and 31.5\%. 
\end{itemize}

Again we verify this attack vectors using PowerWorld. We see that there is no overloading with the attacked load and new generation data during both normal and contingency period. The figure \ref{Fig_Case_3_attacked} shows that when line 7 (bus  4-5) trips there is no overloading among any of the lines.  However, Fig.~ \ref{Fig_Case_3_actual} shows the scenario with the new generations and actual loads, where we see when line 7 trips, line 1, 2, 5, and 6  get overloaded by 2\%, 2\%, 3.75\%, and 31.5\%. We also get the same result from our proposed model.

%%%%%%%%%%%%%%%%%%%%%%%%
\begin{figure*}[t]
    \begin{center}
    %\hspace{-12pt}    
    \subfigure[]
        {
        \label{Eval_2_1}
            \includegraphics[scale=0.5, keepaspectratio=true]{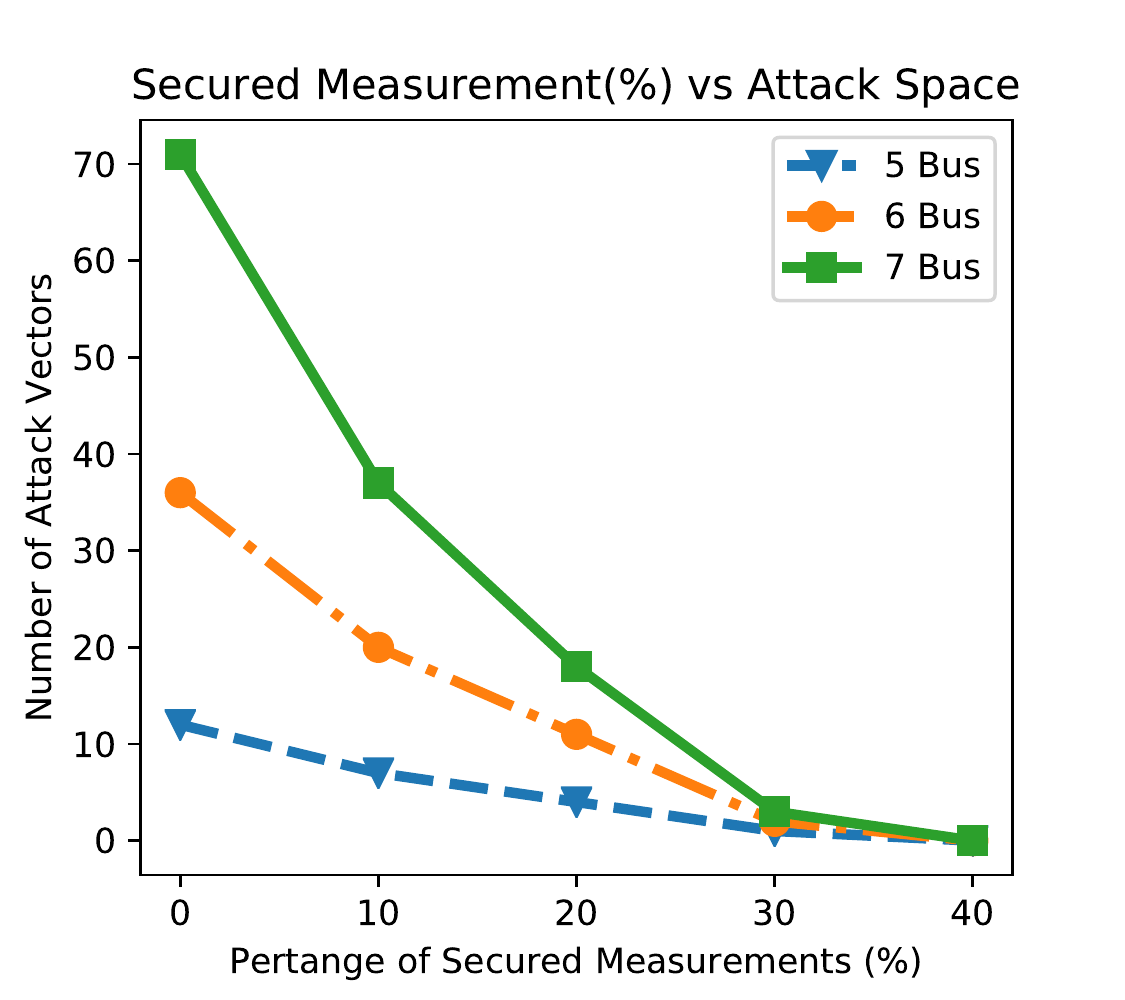}
        }
        \subfigure[]
        {
        \label{Eval_2_2}
            \includegraphics[scale=0.5, keepaspectratio=true]{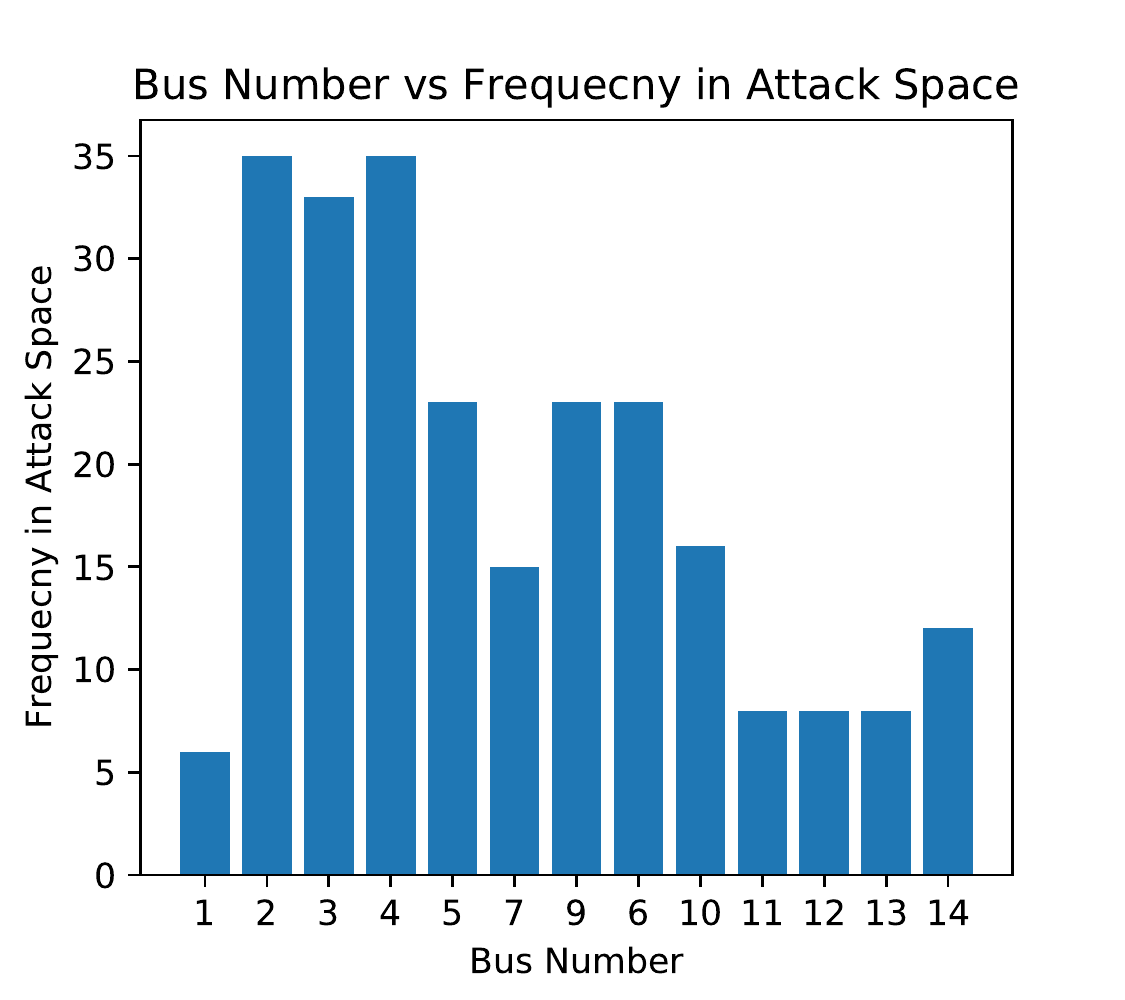}
        }        
        \subfigure[]
        {
        \label{Eval_2_3}
            \includegraphics[scale=0.5, keepaspectratio=true]{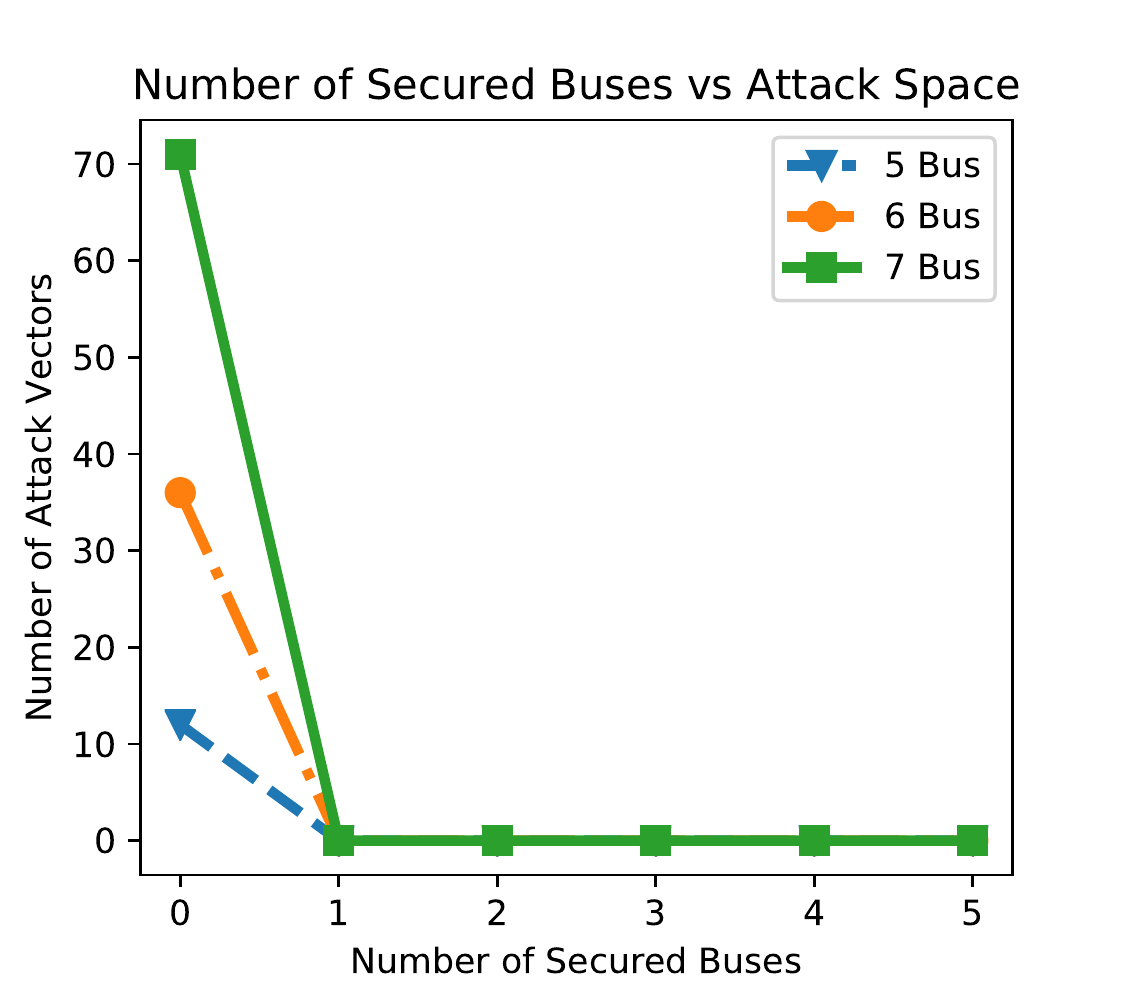}
        }
    \end{center}
    \vspace{-15pt}
    \caption{Model performance on secured measurements with 20\% attacker's capability (a) Impact of randomly secured measurements, (b) bus frequency in attack vectors  and (c) impact of analytically secured measurements.}
    \label{Fig:Eval_2}
\vspace{-12pt}
\end{figure*}
%%%%%%%%%%%%%%%%%%%%%%%%

%%%%%%%%%%%%%%%%%%%%%%%%
\begin{figure*}[t]
    \begin{center}
    %\hspace{-12pt}    
    \subfigure[]
        {
        \label{Eval_3_1}
            \includegraphics[scale=0.5, keepaspectratio=true]{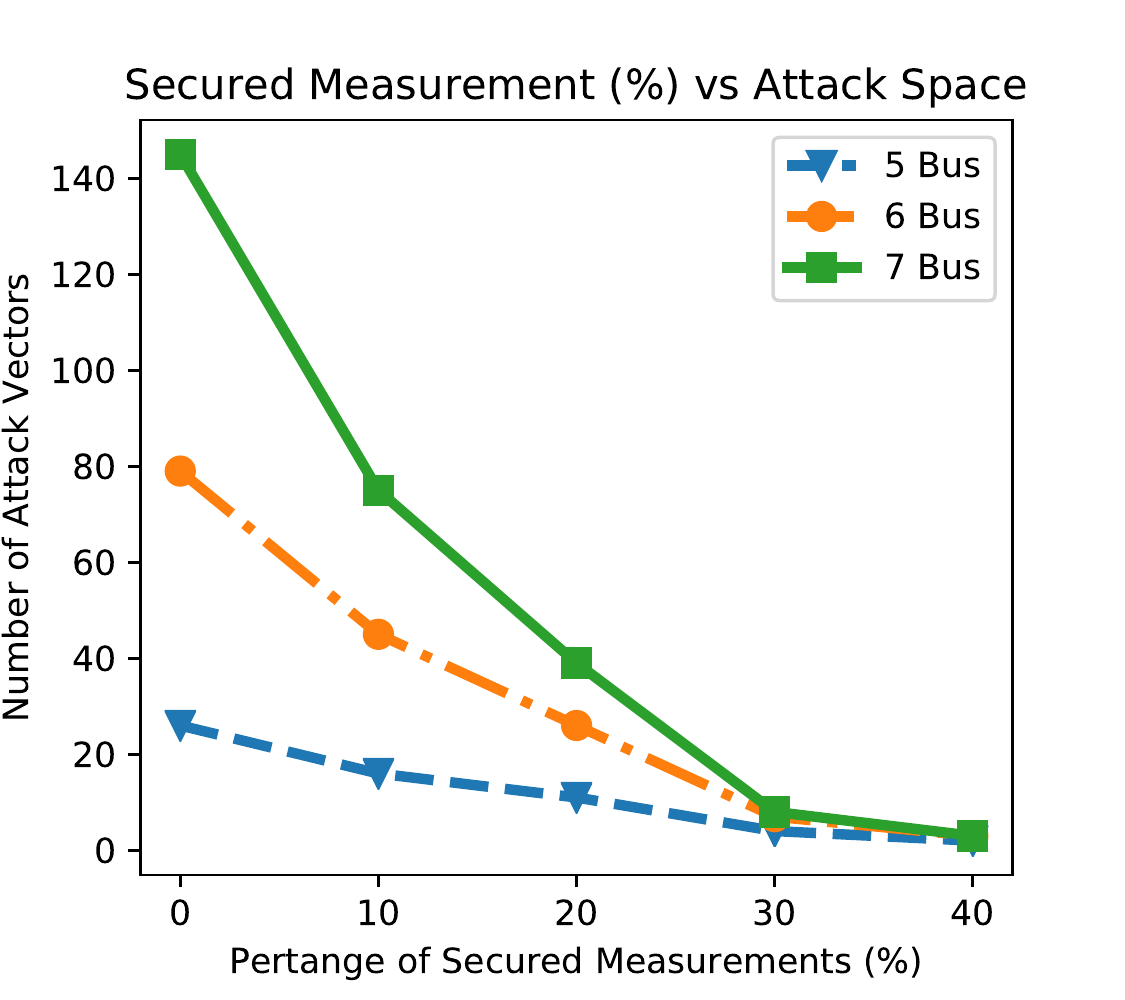}
        }
        \subfigure[]
        {
        \label{Eval_3_2}
            \includegraphics[scale=0.5, keepaspectratio=true]{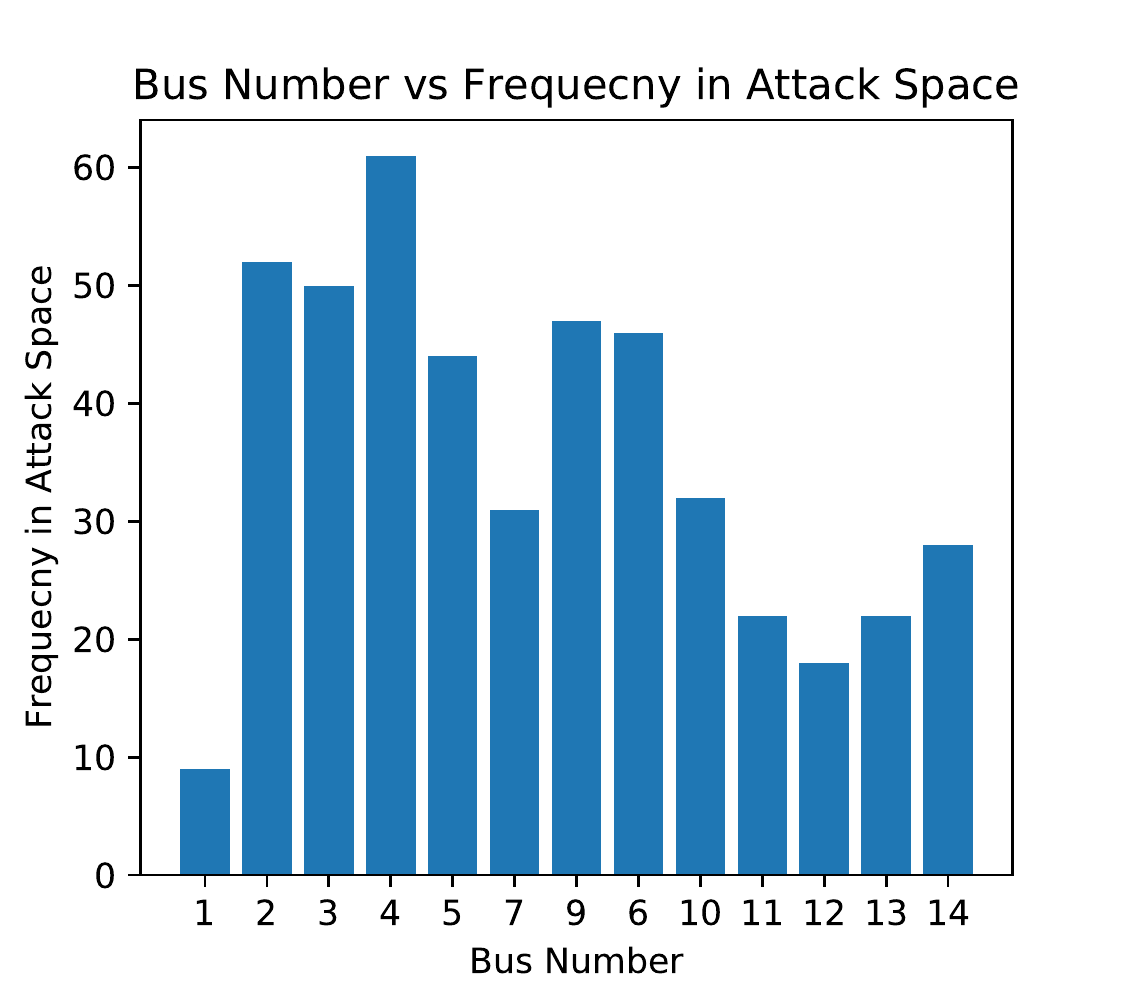}
        }        
        \subfigure[]
        {
        \label{Eval_3_3}
            \includegraphics[scale=0.5, keepaspectratio=true]{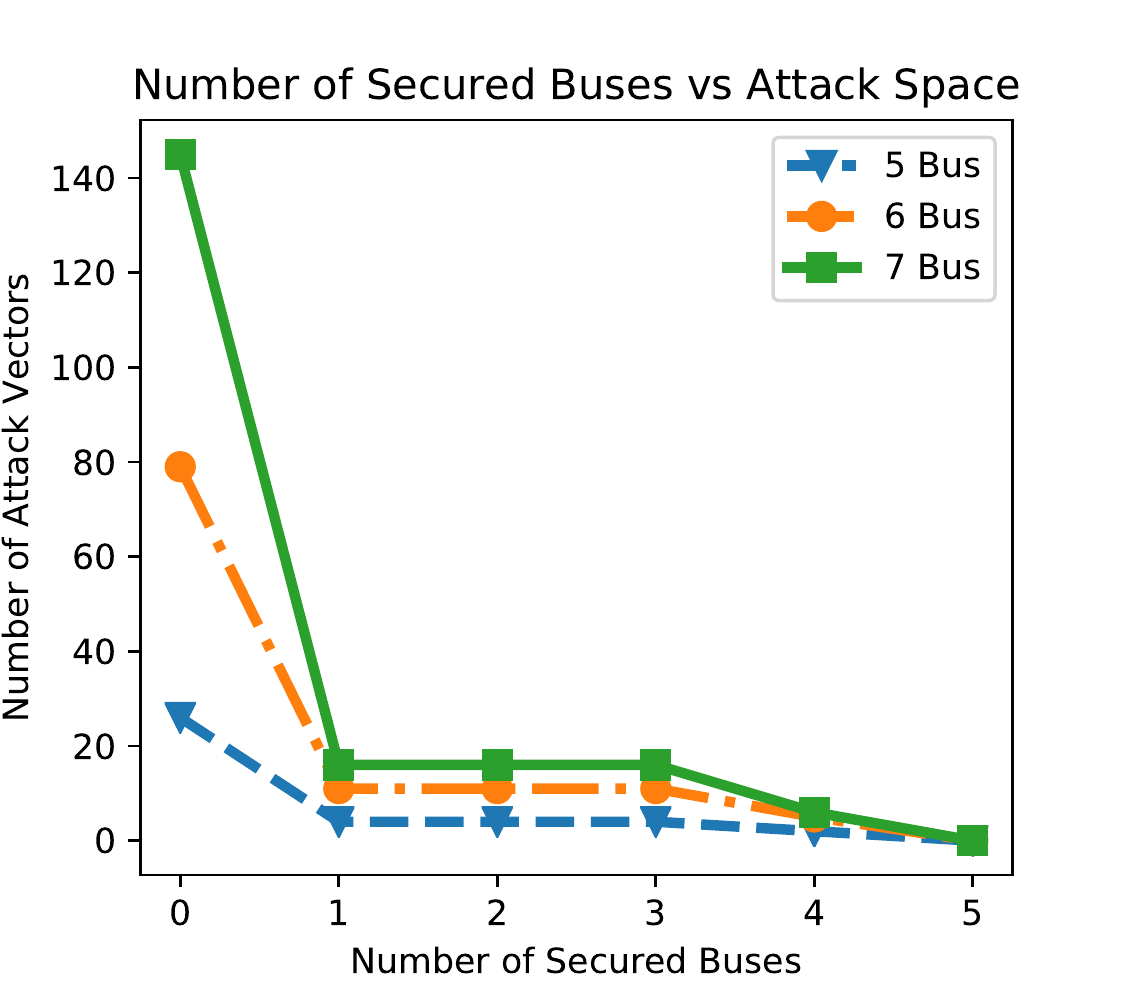}
        }
    \end{center}
    \vspace{-15pt}
    \caption{Model performance on secured measurements with 50\% attacker's capability (a) Impact of randomly secured measurements, (b) bus frequency in attack vectors  and (c) impact of analytically secured measurements.}
    \label{Fig:Eval_3}
\vspace{-12pt}
\end{figure*}
%%%%%%%%%%%%%%%%%%%%%%%%

%%%%%%%%%%%%%%%%%%%%%%%%
\begin{figure*}[t]
    \begin{center}
    %\hspace{-12pt} 
     \subfigure[]
        {
        \label{LineHitmap_1}
            \includegraphics[scale=0.56, keepaspectratio=true]{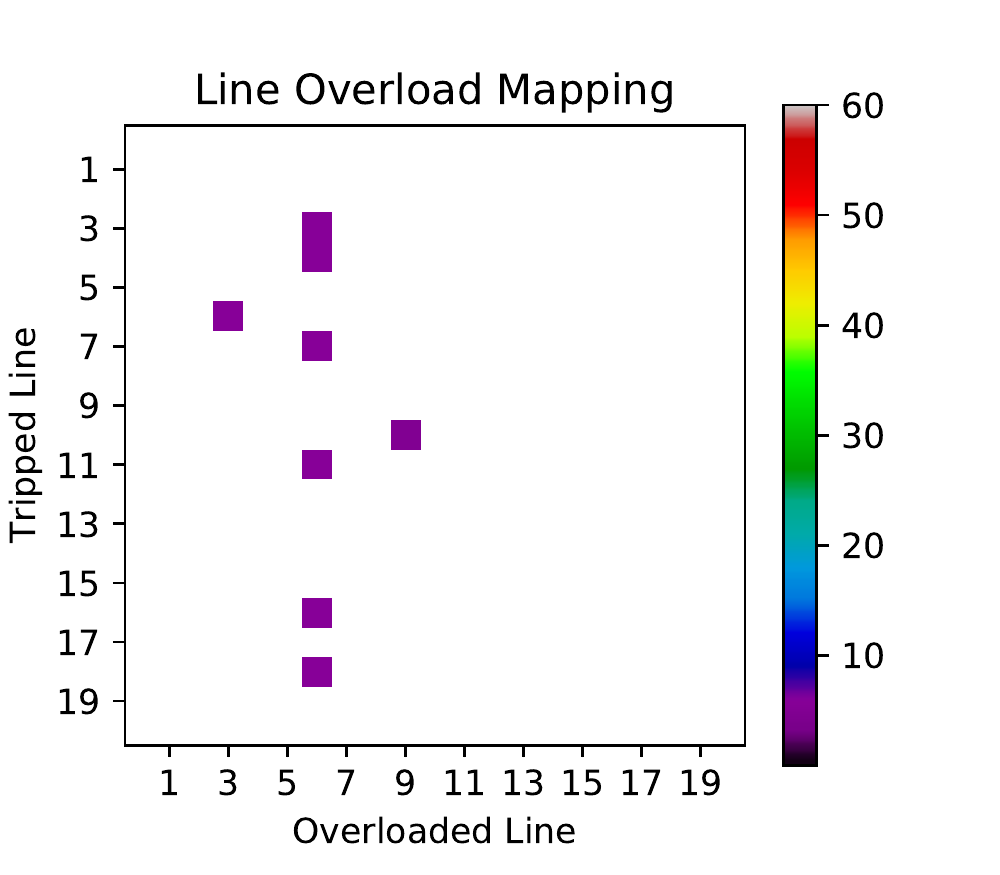}
        }
     \subfigure[]
        {
        \label{LineHitmap_2}
            \includegraphics[scale=0.56, keepaspectratio=true]{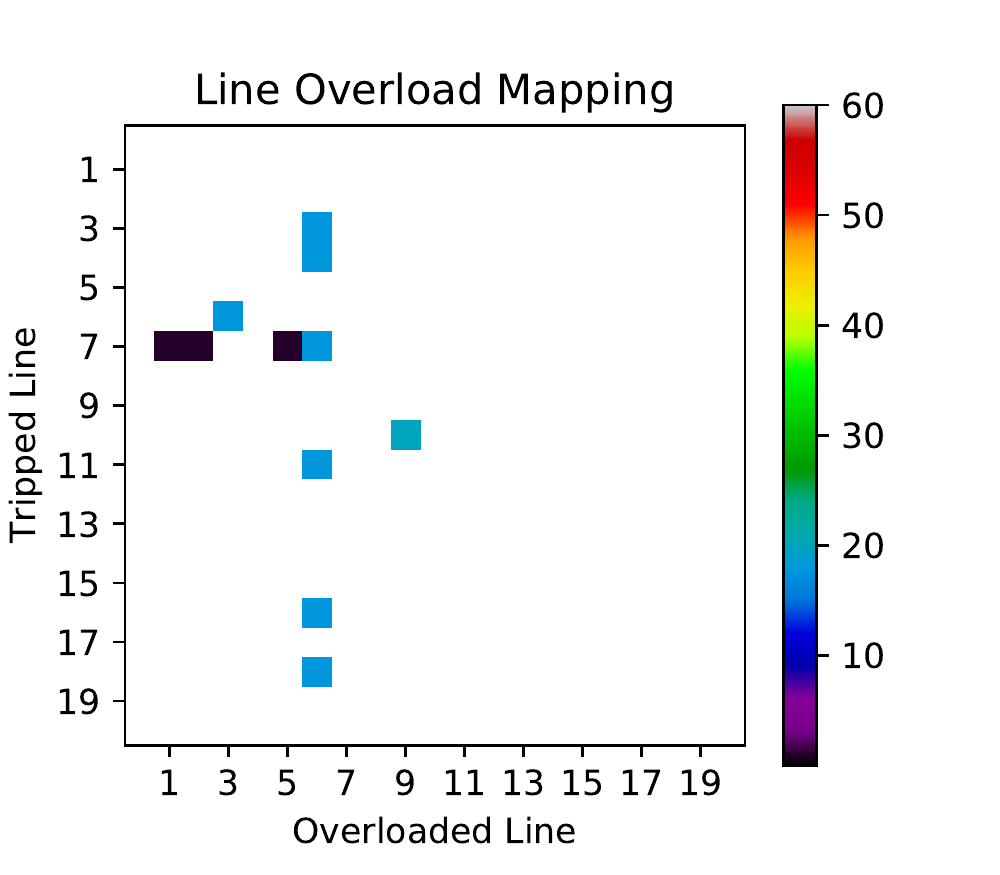}
        }
        \subfigure[]
        {
        \label{LineHitmap_3}
            \includegraphics[scale=0.56, keepaspectratio=true]{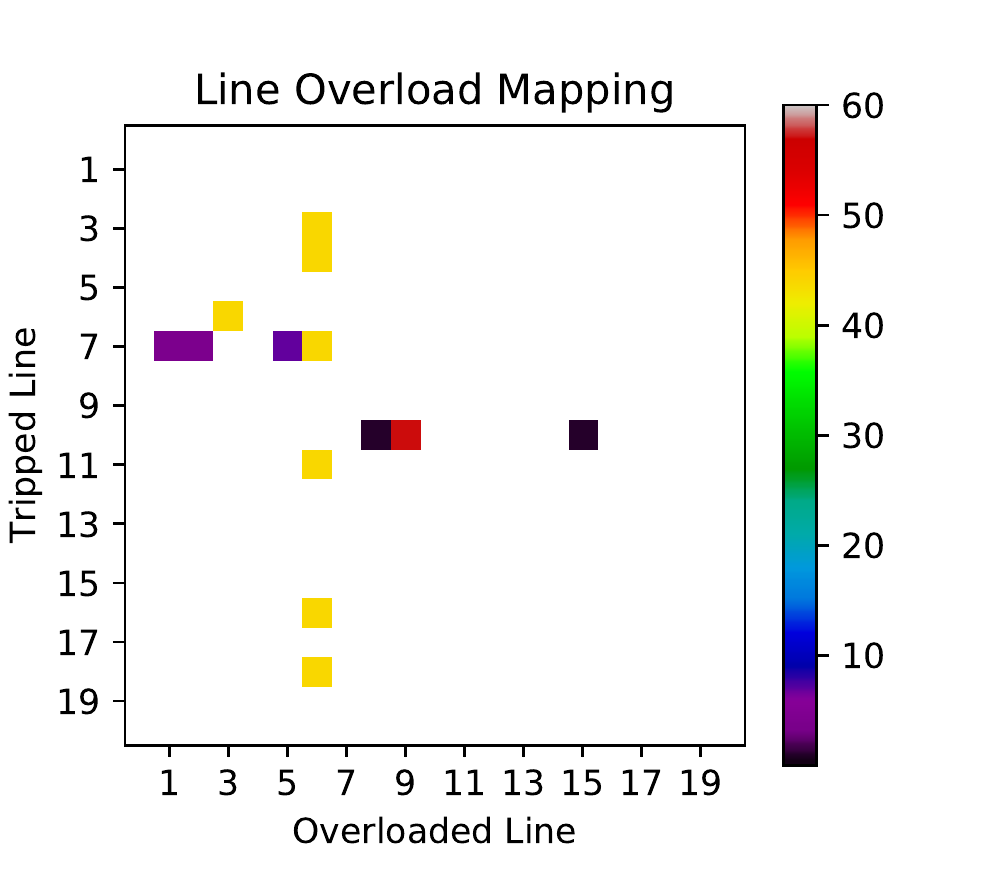}
        }
    \end{center}
    \vspace{-15pt}
    \caption{Heat-map on line overloading scenarios with attacker's capability of attacking measurements at (a) maximum 5 buses, (b) maximum 6 buses, and (c) maximum 7  buses.}
    \label{Fig:LineHitmap}
\vspace{-12pt}
\end{figure*}
%%%%%%%%%%%%%%%%%%%%%%%%

% %%%%%%%%%%%
\section{Evaluation}
\label{Sec:Evaluation}
%Evaluation

The results of the scalability evaluation of the proposed model is presented in this section.
%in terms of time and memory requirements. 
We evaluate the performance of the proposed model with respect to successful UFDI attacks on different
bus states. We use attackability, defined as the number of states which can be attacked (i.e., infected by UFDI attacks) over the total number of states, as the evaluation metric.

%%%%%%%%%%%%%
\subsection{Methodology}
\label{SSec:Methodology}
\vspace{-3pt}

%%%%%%%%%%%%%%%%%%%%%

We evaluate the performance of our proposed model by analyzing the attack space under different scenarios considering the attacker's capabilities, attacker's target, and security measures. To perform this evaluation, we use IEEE 14-bus test system  [10] which consists of 14 buses, 20 transmission lines, and 54 possible measurements as shown in Fig.\ref{Fig_5_Bus}. In our evaluation, we mainly consider three kinds of adversaries with different capabilities to distribute the attack to different buses : (i)  capability to attack a maximum of 5 buses at a time, (ii) capability to attack a maximum of 6 buses at a time, (iii) capability to attack maximum 7 buses at a time. For all of these adversaries, we consider the same resource constraints. An adversary can attack a maximum of 25 measurements at a time.
%%%%%%%%%%%%%%%%%%%%%%%%

\subsection{Evaluation Results and Discussion}
\label{SSec:Scala}%\vspace{-3pt}

\subsubsection{\textbf{Performance with respect to Attacker's Capability}}
Fig.\ref{Eval_1_1} shows the attacker's capability \ie, the percentage of false load that can be injected in buses in different cases considering the attacker's type. Here, we consider the 14-bus test system. The attacker is capable of changing the bus load data from 10\% to 50\% of the original ones. According to the experience, we observe that for the attacker with the capability of attacking a maximum of 5 buses has less number of attack vectors than the attackers with the capability of attacking maximum 6 buses or 7 buses. For the first attacker increasing the attacker's capability does not increase the attack space much; however, for the other two attackers, the size of attack space becomes almost double while increasing the capability from 40\% to 50\%.  %This is expected as with the more attacking capabilities they can fool the system to generate more in the different buses which give more attack vectors.%
However, from 10\% to 40\% of the attacked load, there is no significant increase in the attack space. According to the experiment results, we observe that the attack space is sufficiently large even with changing 10\% of the load.

\subsubsection{Attack Space with respect to the Attacker's Goal}
\noindent \textbf{Percentage of overload of the Lines}
Fig.\ref{Eval_1_2} shows one of the goals of the attackers \ie, the percentage of overloading of the targeted lines,  in different cases with respect to the attacker's capabilities.  The attacker's target is to overload at least 5\% of the total lines (1 line) and have the capability of injecting up-to 30\%  of false load data. Here we evaluate different scenarios with 2\% to 10\% as the overloading amount of those lines.  According to the experiment results, we observe with the increase of the attacker's target to overload more, the number of the attack vectors decreases almost linearly. The results seem logical as overloading lines with 2\% is easier than overloading lines by 10\%. Besides with the target of overloading with a higher percentage, an attacker with the capability of attacking maximum 7 buses has more attack vectors than others. 

\noindent \textbf{percentage of overloaded lines}
Figure \ref{Eval_1_3} shows one of the goals of the attackers \ie, the percentage of the overloaded lines in different cases with respect to the attacker's capabilities. The attacker's target is to overload lines by at least 2\% of the rated capacity with injecting up-to 50\% of false load data in the measurements. Here, we evaluate different scenarios with 10\% to 25\% of the lines (1 line to 4 lines) to be overloaded after any contingency.  According to the experiment results, we observe that with the increase in attacker's target to overload lines, the number of attack vectors decreases rapidly. Satisfying constraints to overload only 1 line is much easier for the model than to overload more lines. The attack space becomes too small when the attack target is to overload 2 to 4 lines. However, overloading 25\% of the lines (5 lines) is not possible with the current attack attributes. 

\subsubsection{Securing the Measurements}
Securing the measurements limits the attacker capability to launch an FDI attack to the power system. As to bypass the BDD algorithm, the attacker may need to change multiple measurements which may not be possible if one or some of them are secured. In the following sections, the impact of the secured measurements is discussed. \par 

\noindent\textbf{Impact of Randomly Secured Measurements}: 
To increase the resiliency of the system, the measurements must have data integrity and need to be secured from all kinds of cyberattacks. Thus, we study the impact of the percentage of secured measurements on the attack space of this example. We explored the attack space by changing the percentage of secured measurements from 0\% to 40\% of the total measurements. Again, we consider two cases where the attacker has the capability to inject 20\% and 50\% of false data, respectively. The target is to overload at least 5\% of the lines with minimum 2\% overloading. Fig.\ref{Eval_1_1} and Fig.\ref{Eval_2_1} show these two cases when the measurements are selected and secured randomly. From both of the figures, it is observed that securing the measurements reduces the attack space almost linearly. However, to make the system completely secured from any UFDI attack, for both of the cases, at least 40\% of the measurements need to be secured when they are chosen randomly. 

\noindent\textbf{Frequency of Compromised Buses}:
Due to different configurations, capabilities and connectivity of the buses, attacking the measurements located at some specific buses gives comparatively more attack vectors than others. To explore this relationship, we analyzed the attack vectors resulted from the two studies as discussed in the earlier section where the attackers can inject up to 20\% and 50\% of the load data respectively to overload at least 5\% of the total lines by 2\% of their respected rated capacity. Thus, analyzing all the attack vectors, we calculate the recurrence of each measurement in the attack vectors. From Fig.\ref{Eval_1_2} and Fig.\ref{Eval_2_2}, it is clear that measurements at some specific buses are more common. From both of the bar charts, we see that compromising the measurements at Bus 2, 3, 4, 5, 6 and 9 are giving more attack vectors than other do. From these results, it is clear that buses with both generator and load are more frequent in the attack space than others. Besides, the neighboring load buses connected to these buses are also more common in the attack scenarios. Hence, as the attacker's target is not the increase in the generation cost, the buses with the lower generation cost are more frequent in the attacks scenarios. 

\noindent\textbf{Impact of Analytically Secured Measurements }:
In this part, the buses with higher frequencies are secured first and the impact is studied under the same attack attributes. Thus, from the sorted list of the highly affected buses, we secure 1, 2, 3, and 4 most common buses in the attacks respectively and then analyze the attack vectors again for each of the cases. For the first case (with 20\% data injection), only securing the measurements at bus 4, which is the most frequent one, makes all the attack vectors invalid. Thus, with this approach securing only 7\% of the total measurements makes the system robust from all the UFDI attacks on SCOPF, whereas 40\% was needed if it was done randomly. In the second case where the attacker is more resourceful with 50\% data injection capability, there are more attack vectors. Again, securing only the most frequent buses reduces the attack spaces drastically. Fig.\ref{Eval_3_3} shows securing only the measurements at bus 4 (7\% of total measurements) reduces the attacks space by almost 90\% where 30\% is needed when it is done in arbitrary. 

\subsubsection{Line Overload Heatmap}:
Fig.\ref{Fig:LineHitmap} shows the heat map of the line overloading for 3 different cases. In the x-axis, it shows the lines that need to trip to make others overloaded and in the y-axis, it shows the lines get overloaded. Fig.\ref{LineHitmap_1} shows the heatmap when the attacker has the capability of attacking measurements distributed at a maximum of 5 buses. It is clear from the figure that for most of the attack vectors line 6 gets overloaded when the other lines trip; there are also some other scenarios when line 3 and 9 get overloaded. Fig.\ref{LineHitmap_2} shows the scenarios of line overloading when the attacker can attack up to 6 buses at a time. Now, it gives additional scenarios where line 1, 2 and 5 get overloaded as well. Finally, Fig. \ref{LineHitmap_3} represents the scenarios attacking measurements distributed at up to 7 buses. With this capability, now the attacker can overload line 1, 2, 3, 5, 6, 8, 9 and also 15. All these lines except 6, get overloaded only when line 6, 7 or 10 trips which are mainly connected to the generation bus. Thus, making these lines physically more robust reduces the probability of the attacker success.

\section{Related Work}
\label{Sec:Related}
%\textbf{Related Work from TDSC}
%\input{Related_Work_TDSC.tex}

Cyber attacks on power grids have been considered in literature over the last few years~\cite{Salmeron04, Ten08}. Nowadays, Undetected False Data Injection (UFDI) attacks are one of the most researched and continuing issues on the cybersecurity analysis of power systems. It mainly considers the concept of stealthy attacks against the system control loop. The idea of such stealthy attacks is first introduced in~\cite{Liu09}, and continued succeeding in~\cite{Liu11}. The authors represented different cases to explain the attack. One of them was the attacker's limitation of the access to resources and sensors/meters to alter them. They considered the adversary has comprehensive knowledge about the network and the targets are arbitrary and specific.  In the usual circumstance, finding the attack vector for a closed-loop system is an NP-complete problem. Therefore, the authors performed a few heuristic procedures to obtain attack vectors.
%In~\cite{Liu11}, the authors extended UFDI attacks considering different scenarios, such as limited access and limited resources, etc., assuming that the adversary has complete knowledge of the grid. 
%In the general case, the attack vector computation problem is shown as NP-complete. Therefore, the authors presented few heuristic approaches that can find attack vectors.
%

Vukovic et al. introduced several security metrics to calculate the weight of individual buses and the expense of compromising different measurements acknowledging the vulnerability of the system connectivity~\cite{Vukovic11}. Kin Sou et al. proposed that an $l_1$ relaxation-based method gives an accurate optimal solution of the attack vector generation problem~\cite{Sou13}. 

%UFDI attacks with incomplete or partial information are discussed by the works presented in~\cite{Teixeira10, Ashfaq12}.
% 

UFDI attacks with partial or inadequate knowledge of the power system equipment properties are presented in~\cite{Teixeira10, Ashfaq12}. It is also shown in~\cite{Esmalifalak11} that an adversary can launch UFDI attacks despite not knowing the topology. The authors estimated the linear composition of the topology from the measurements to initialize a UFDI attack based on that. The work \cite{Esmalifalak13} shows the competition between an intruder and a defender concerning the effects of UFDI attacks on energy markets.

In \cite{Kim13}, the algebraic forms of undetected topology attacks in power grids are introduced. The idea of anonymous attacks is proposed in~\cite{Qin12}, where the grid operator might be able to detect the presence of bad data, but cannot recognize the compromised sensor accurately. In ~\cite{Yuan11}, the authors introduced load redistribution attacks which were discussed later for situations where the attacker has inadequate knowledge ~\cite{Liu14}. 
Rahman et al.  introduced a formal model which verify stealthy attacks comprehensively on state estimation considering different attack attributes concurrently~\cite{RahmanDSN14}. Later in ~\cite{RahmanICCPS14}, they assessed the effect of stealthy attacks on the economic performance of the power system, considering the interrelation between SE and OPF and topology poisoning-based stealthy attacks in~\cite{RahmanICDCS14}. In~\cite{rahman2018impact} the authors investigated the feasibility and economic influence of stealthy attacks using Simulink and formal model. 

 In formal modeling, few works tried to use Simulink based design as well. In~\cite{Roy2011}, the authors suggest a verifier for a contract-based scheme, which is created using simulink that certifies the accuracy of the Simulink model. In~\cite{Araiza14}, the authors used Simulink based modeling to verify the control system properties. They used Why3, a platform for deductive program verification. In~\cite{Bobba10}, Bobba et al. proposed that defending a set of measurements which assure observability can detect UFDI attacks.
In~\cite{Kim11}, the authors suggested a greedy suboptimal algorithm that can determine a measurement subset, which is immune to false data injection. 

Recently, Kang et al.~\cite{kang2018false} proposed an FDI attack on contingency analysis of power system, where the target was to mislead the operational cost only. However, if the system runs SCOPF on false data, both physical and economic impacts are there. The authors did not consider any consequences as physical damage in their work. In~\cite{chu2019vulnerability}, Chu at el. studied the vulnerability of power system by targeting a specific line when another line trips. They  only considered few specific probable cases and the false data is injected to the state estimator instead of the true measurements ignoring the security/accessibility of the measurements.  Chu at el. studied the vulnerability of power system by targeting a specific line when another line trips~\cite{chu2019vulnerability}. The proposed framework considers only few probable cases and the false data is injected directly to the state estimator instead of the true measurements. They ignored the impacts of the security/accessibility of the measurements on the attack vector. 

%\textbf{We need to write some motivational lines here :} \par
Unlike the above mentioned works, we propose a comprehensive formal framework to synthesize the stealthy attack vectors that can compromise the solution by CA, providing a false notion of security in contingencies. This synthesis model considers a complete attack path, from altering the measurements to their impact path down to CA, and can automatically identify the critical threat space.

%It is worth mentioning that, to our best knowledge, our work is the first of its kind in proposing, modeling and analyzing cyber-vulnerabilities via a formal satisfiability framework, especially on modules that depend on the state-estimator.

%%%%%%%%%%%
%%%%%%%%%%%
\section{Conclusion}
\label{Sec:Conclusion}
%\vspace{-6pt}

In this work we have shown that impact of stealthy attacks on SE propagates to SCOPF, which causes incorrect dispatch decisions, leading to transmission line overloading in the case of contingencies. In such cases, the grid can face the tripping of more transmission lines, and thus power outage.
%Our work shows that attacks on the state estimator can be exploited and further strengthened to induce vulnerabilities in the Optimal Power Flow (SCOPF) module. 
We have presented a formal model that allows us to find such SCOPF-affected threats and its causes (attack vectors). We have analyzed the feasibility of these attacks with respect to IEEE bus systems considering different attack environments. The results shows that our proposed tool can explore the potential attack space efficiently. We have also leveraged PowerWorld to verify the proposed model.
The proposed formal model would be useful to provably identify and analyze the resiliency of a grid's correct and efficient operation against cyber attacks, which will assist in applying necessary security measures for a secure and dependable system.

%%%%%%%%%%%%%%%%%
%%%%%%%%%%%%%%%%%
\bibliographystyle{unsrt}
\bibliography{SG_SCOPF_References}

\appendix

%%%%%%%%%%%%%%%%%%%%%%%%%%%%%%%%

\subsection{DC Power Flow Model}
\label{Appendix:DCPowerFlowModel}

%The most common and important calculation for a power system is the AC power flow analysis. However, an AC power flow analysis is computationally expensive as a grid with $n$ nodes needs to solve $2n$ non-linear equations through iteration. The high level of accuracy with a detailed result provided by the AC power flow model does not overcome the high computational expense. Therefore, a linearized model of the system is used to speed up the computation which is called  DC power flow model~\cite{purchala2005usefulness}. To make the model less expensive, some assumptions are made which speed up the computation with the trade of accuracy. However, the result of the DC power model is accurate enough the load flow calculation.  
In DC model, the first assumption is to consider the transmission line resistances ($R_L$) insignificant compared to the line reactances. %($X_L$): $(R_L << X_L)$. 
This means lines are considered purely inductive. 
Considering $R_L$ as resistance, $G_L$ as conductance, $X_L$ as reactance, $B_L$ as susceptance, $Z_L$ as impedance and $Y_L$ as the admittance of a transmission line $L$:
\begin{equation}
    \begin{split}
        & G_L=\frac{R_L}{R_L^2+X_L^2}~\approx~0\\
        & B_L=\frac{-X_L}{R_L^2+X_L^2}~\approx~-\frac{1}{X_L}\\
        & Z_L=\approx j.X_L ~~\And~~ Y_L\approx~j.B_L
    \end{split}
\end{equation}

The second assumption is that the voltage of the buses is considered as fixed at 1.0 per unit (pu). The only variables of the system are the states of the buses which are the phase angles and those can be calculated from the different sensor measurements using state estimation. Then, considering all the assumptions, the phasor ($\theta_j$) of the voltage ($V_j$) at bus $j$ is considered as:
$V_j \approx 1\angle\theta_j$.

Line admittance is the reciprocal of the line reactance. The phase differences between two connected buses are small enough to results in the linearization of $sine$ and $cosine$ functions. Thus:
\begin{equation}
    \begin{split}
&\sin (\theta_i-\theta_j) \approx \theta_i-\theta_j \\
& \cos(\theta_i-\theta_j) \approx 1\\
    \end{split}
\end{equation}

Let the admittance of the line connecting buses $i$ and $j$ is $Y_{ij}$, thus the power flow through the transmission line these buses is calculated by: $P_{ij}=Y_{ij}(\theta_i-\theta_j)$ where $\theta_{i}$ and $\theta_{j}$ are the phase angles of bus $i$ and $j$ respectively. Considering the power flow constraint, the algebraic sum of the power flow incoming to a bus should be zero. Assuming the system has $n$ buses, the constraint can be modeled using a linear system equation  $\textbf{B}[\theta]=[\textbf{P}]$, where $[\textbf{B}]$ is a $n$-1 dimensional square matrix, $\textbf{P}$ is a matrix with the same size of $\textbf{B}$ where its elements show the net power demand at a bus, and $[\theta]$ is a column vector of unknown phases corresponding to the bus voltage phasors. There is a reference bus for which the phase angle $\theta_i=0$. 
The model solves the unknown phase angles, given the power amounts (i.e., generation and/or load) at all the buses and the line reactances. This linear model gives the base for DC state estimation described in the following section.

%%%%%%%%%%%%%%%%%%%%%%%%%%%%%%%%%%%%%%%
\subsection{Line Outage Distribution Factors (LODFs)}
\label{Appendix:LODF}
Line Outage Distribution Factors (LODFs) represent the sensitivity of a power system during a contingency condition i.e. line outage~\cite{guo2009direct}. Power system is a complex network and all of the elements have their own operating limit. Thus, if power flow through any line exceeds its rating, circuit breaker will trip the line to protect the rest of the system. On other hand, there might be also some natural or accidental events that may also be the reason behind the contingency. During the outage, pre-outage power flowing through the affected line is distributed to the other lines with respect to a sensitivity factor matrix that is called Line Outage Distribution Factors (LODFs). Each elements of the matrix represent if one line is tripped what percent of the power of the affected line will be shared by the another line. Thus, LODFs are used to calculate the linear impact of contingencies in power system simulator. Therefore, in a power system during the outage of $line_2$, post contingency power flow of $line_1$ is

$$Power_{line_1}^{post}= Power_{line_1}^{pre} + LODF_{line_1,line_2} \times Power_{line_2}^{pre}$$
where $Power_{line_1}^{pre}$ and $Power_{line_2}^{pre}$ are the pre-outage power flows of $line_1$  and $line_2$ respectively and $LODF_{line_1,line_2}$ is LODF of $line_1$ with respect to the outage of $line_2$.  \par

In our work to calculate the LODF matrix of a power system~\cite{al2016simulation}, 
firstly we calculate the $Y_{bus}$ matrix as
\begin{gather}
Y_{bus}=\begin{bmatrix}
   y_{11} &  y_{12} & ....  & y_{1n}\\  
   y_{21} & y_{22}  & .... & y_{2n}\\
   ....   & ....  & .... & ....\\
   ....   & ....  & .... & ....\\
   y_{n1} & y_{n2} & .... & y_{nn}\\
   \end{bmatrix}
\end{gather}

\begin{equation}
    \begin{split}
        &y_{ij}=-\frac{1}{z_{ij}}\\
 &y_{ii}=\sum_{{i=1}~ {i\ne j}}^{n} y_{ij}
    \end{split}
\end{equation}

where $z_{ij}$ is the impedance of the line between bus i and j. We consider bus 1 as the slack bus and thus eliminating the $1^{st}$ row and $1^{st}$ column from the $Y_{bus}$ matrix, we get

\begin{gather}
Y_{eliminate}=\begin{bmatrix}
   y_{22} & y_{23}  & .... & y_{2n}\\
    y_{32} &  y_{33} & ....  & y_{1n}\\  
   ....   & ....  & .... & ....\\
   ....   & ....  & .... & ....\\
   y_{n2} & y_{n3} & .... & y_{nn}\\
   \end{bmatrix}
\end{gather}

The sensitivity matrix is

\begin{gather}
X=\begin{bmatrix}
   0 & 0  \\
   0 & Y_{eliminate}^{-1} \\
   \end{bmatrix}
\end{gather}
 
 Finally, the line outage distribution factor (LODF) of line $line_1$ due to the outage of the line $line_2$  can be calculated by the following equation

 \begin{equation}
     LODF_{(line_1,line_2)}=\frac{{\frac{z_{ij}}{z_{kl}}}(X_{il}-X_{jl}-X_{ik}+X_{jk})}{z_{kl}-(X_{ii}+X_{jj}-2X_{ij}}
 \end{equation}

%%%%%%%%%%%%%%%%%%%%%%%%%%%%%%%%
\subsection{Case Study Input}
\label{Appendix:CaseStudyInput}

Table~\ref{Tab_Input} shows the primary input data (partial) for our example case studies.

%%%%%%%%
\begin{table}[t]
%\vspace{-12pt}
\caption{Input (Partial) for Case Study 1} 
\label{Tab_Input} 
\vspace{-6pt}
\centering
\begin{tabular}
{p{3.2in}}
\hline 

\vspace{0.01in}
\# Topology (Line) Information \\
\# (line no, from bus, to bus, admittance, line capacity, knowledge?, in true topology?, in core?, secured?) \\
1    1    2    16.90    0.65\\
2    1    5    4.48    0.51\\
3    2    3    5.05    0.15\\
.........................................\\
.........................................\\
19    12    13    5.00    0.35\\
20    14    13    2.87    0.55\\

%\ldots    \ldots    \ldots    \ldots    \\
\vspace{0.01in}

\# Bus Types (bus no, is generator?, is load?) \\
1    1    0\\
2    1    1\\
3    1    1\\
.........\\
.........\\
13    0    1\\
14    0    1\\
\vspace{0.01in}

\# Generator Information (bus no, max generation, min generation, cost coefficient) \\
1    0.80    0.10    20    220\\
2    0.80    0.10    20    150\\
...........................\\
6    1.80    0.10    20    100\\
8    0.40    0.10    20    250\\
\vspace{0.01in}

\# Load Information (bus no, existing load, max load, min load) \\
2    0.217    0.60    0.01\\    
3    0.80    1.50    0.01\\    
........................\\
........................\\
13    0.135    0.35    0.01\\    
14    0.149    0.35    0.01\\
\vspace{0.01in}

\# Measurement Information \\
\# (measurement no, measurement taken?, secured?, can attacker alter?) \\
1    1    1    0\\
2    1    0    1\\
3    1    0    1\\
.............\\
.............\\
.............\\
53    1    1    1\\
54    1    0    1\\

%\ldots    \ldots    \ldots    \ldots    \\
\vspace{0.01in}

\# Cost Constraint\\
382\\
\vspace{0.01in}

\# Attacker's Resource Limitation (measurements, buses) \\
20    3\\
\vspace{0.01in}

\# Maximum percent of delta load\\
20\\
\vspace{0.01in}

\# \% of minimum Overloading amount, \% of lines to be overloaded\\
5 5\\
\vspace{0.01in}

\\\hline
\end{tabular}
\normalsize
\vspace{-15pt}
\end{table}
%%%%%%

\end{document}